\documentclass[journal]{IEEEtran}
\usepackage{amsmath,amssymb,amsfonts}
\usepackage{algorithmic}
\usepackage{algorithm}
\usepackage{array}
\usepackage{textcomp}
\usepackage{stfloats}
\usepackage{url}
\usepackage{verbatim}
\usepackage{graphicx}
\usepackage{multirow}
\usepackage{subfigure}
\usepackage{threeparttable}
\usepackage{float}
\usepackage{booktabs}
\usepackage{notoccite}
\usepackage{longtable}
\usepackage{makecell}
\usepackage{lineno}

\begin{document}

\title{Integrating Language-Image Prior into EEG Decoding for Cross-Task Zero-Calibration RSVP-BCI}

\author{Xujin Li, Wei Wei, Shuang Qiu, Xinyi Zhang, Fu Li, and Huiguang He, \IEEEmembership{Senior~Member,~IEEE}% <-this % stops a space

\IEEEcompsocitemizethanks{\IEEEcompsocthanksitem X. Li, W. Wei, S. Qiu, X. Zhang and H. He are with the Key Laboratory of Brain Cognition and Brain-inspired Intelligence Technology, Institute of Automation, Chinese Academy of Sciences, Beijing
100190, China. X. Li, X. Zhang, and H. He are also with the School of Future Technology, University of Chinese Academy of Sciences (UCAS), Beijing 100049, China (E-mail: lixujin2021@ia.ac.cn, weiwei2018@ia.ac.cn, shuang.qiu@ia.ac.cn, zhangxinyi2020@ia.ac.cn, huiguang.he@ia.ac.cn).}

\IEEEcompsocitemizethanks{\IEEEcompsocthanksitem F. Li is with the Key Laboratory of Intelligent Perception and Image Understanding of Ministry of Education, the School of Artificial Intelligence, Xidian University, Xi’an, 710071, China. (E-mail: fuli@mail.xidian.edu.cn).}

\thanks{(Corresponding authors: Huiguang He, Shuang Qiu.)}
\thanks{The first two authors contributed equally to this work.}
\thanks{Manuscript received April 19, 2005; revised August 26, 2015.}}

\maketitle

\begin{abstract}
    Rapid Serial Visual Presentation (RSVP)-based Brain-Computer Interface (BCI) is an effective technology used for information detection by detecting Event-Related Potentials (ERPs). The current RSVP decoding methods can perform well in decoding EEG signals within a single RSVP task, but their decoding performance significantly decreases when directly applied to different RSVP tasks without calibration data from the new tasks. This limits the rapid and efficient deployment of RSVP-BCI systems for detecting different categories of targets in various scenarios. To overcome this limitation, this study aims to enhance the cross-task zero-calibration RSVP decoding performance. First, we design three distinct RSVP tasks for target image retrieval and build an open-source dataset containing EEG signals and corresponding stimulus images. Then we propose an EEG with Language-Image Prior fusion Transformer (ELIPformer) for cross-task zero-calibration RSVP decoding. Specifically, we propose a prompt encoder based on the language-image pre-trained model to extract language-image features from task-specific prompts and stimulus images as prior knowledge for enhancing EEG decoding. A cross bidirectional attention mechanism is also adopted to facilitate the effective feature fusion and alignment between the EEG and language-image features. Extensive experiments demonstrate that the proposed model achieves superior performance in cross-task zero-calibration RSVP decoding, which promotes the RSVP-BCI system from research to practical application.
\end{abstract}

\begin{IEEEkeywords}
    Brain-Computer Interface (BCI), Rapid Serial Visual Presentation (RSVP), Cross-task zero-calibration, Transformer.
\end{IEEEkeywords}

\section{Introduction}
    \IEEEPARstart{T}{he} Brain-Computer Interface (BCI) system enables direct communication between the brain and external devices, with applications in communication, control, and rehabilitation \cite{wolpaw2002brain}. Rapid Serial Visual Presentation (RSVP)-based BCIs have received substantial attention for their ability to enhance human-computer interaction and increase human capabilities \cite{lees2018review}. These systems have been employed in various applications, such as target image retrieval \cite{marathe2015improved,wei2020reducing}, speller systems \cite{acqualagna2013gaze}, face recognition \cite{chen2024eeg}, and anti-deception \cite{wu2018anti}. Among them, target image retrieval is the most extensive and typical application of an RSVP-BCI system.
    
    In the RSVP paradigm, sequences of images that include target images among numerous nontarget images are displayed at high speed (typically 5–10 Hz) in the same spatial location \cite{lees2018review,qiu2023survey}. Subjects are instructed to identify task-specific target images specified in advance, and the rare target images evoke Event-Related Potentials (ERPs) that contain the P300 potential in EEG signals \cite{polich2007updating}. The P300 potential is a prominent wave evoked by rare target stimuli, typically peaking 300–500 ms after the onset of a target stimulus, such as a visual or auditory stimulus \cite{zhang2021evaluation}. Detecting P300 potentials facilitates the classification of target images within image streams.

    Previous studies have developed conventional machine learning methods for P300 detection in RSVP tasks \cite{HDCA,MDRM,rivet2009xdawn}. Over the past decade, with advancements in deep learning, numerous deep learning approaches have been introduced to improve RSVP decoding performance \cite{MCNN,PLNet,PPNN,ji2024novel}. Due to individual differences in brain activity \cite{morioka2015learning}, these methods require training and testing on EEG data collected from the same subjects performing the same RSVP task to ensure reliable performance, which is also called subject-dependent decoding. Applying RSVP-BCI systems to new subjects requires a time-consuming calibration process to collect training data from new subjects and train subject-specific models. This process restricts the application of RSVP-BCI in practice. To avoid the time-consuming calibration procedure, researchers have developed zero-calibration decoding, also known as subject-independent decoding, by utilizing existing labeled data to train the model that can be applied to new subjects without calibration. Zero-calibration decoding primarily focuses on single RSVP tasks, where the system operates in the same scenario and retrieves the same target category. Several zero-calibration methods have demonstrated performance comparable to methods in subject-dependent decoding on new subjects \cite{LeeCNN,wei2022erp,li2022tff}. Therefore, RSVP-BCI systems employing zero-calibration decoding methods can directly decode EEG signals from new subjects performing the same RSVP task without calibration, which can save preparation time and facilitate rapid deployment in practical applications.

    \begin{figure*}[!htbp] 
        \centering
        \subfigure[]{
    	\includegraphics[width=0.4\linewidth]{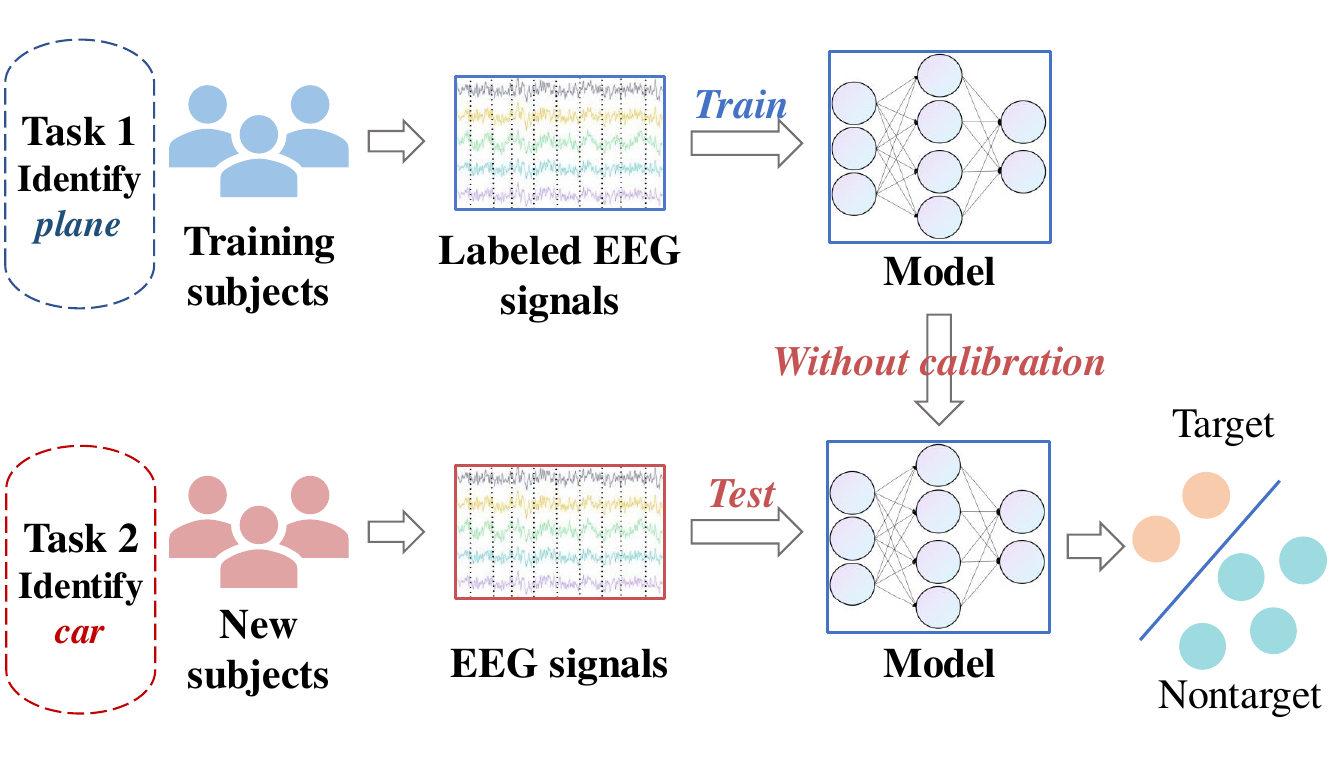}
    	}\hspace{0.12cm}
        \subfigure[]{
    	\includegraphics[width=0.56\linewidth]{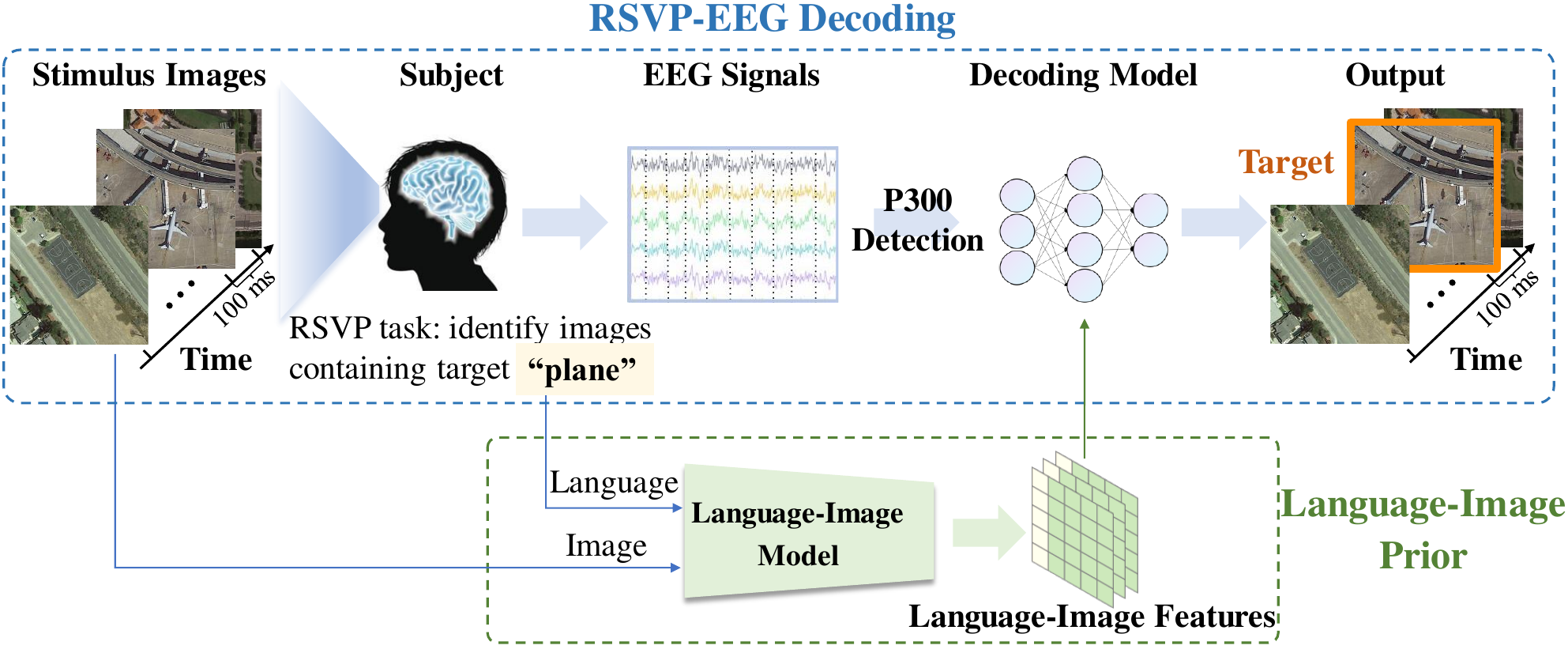}
    	}
        \caption{The diagram of (a) cross-task zero-calibration decoding and (b) RSVP-EEG decoding and integrating language-image prior into EEG decoding. (a) The RSVP decoding model is trained on EEG signals from training subjects performing existing RSVP tasks (i.e. Task plane), and can be directly used to efficiently classify EEG signals from new subjects performing new RSVP tasks (i.e. Task car). (b) The image sequence is presented to the subject at a high rate (e.g., 10 Hz) while the subject's EEG signals are recorded. The decoding model identifies ERPs in EEG signals to classify corresponding stimulus images as target images. Language-image features are extracted from task-specific prompts and stimulus images using language-image models to enhance RSVP decoding.}
        \label{cross-task RSVP-BCI}
    \end{figure*}

    However, while these methods perform well on single RSVP tasks where the RSVP-BCI operates in the same scenario with a fixed target category, their performance declines significantly in cross-task zero-calibration decoding \cite{wei2022erp,waytowich2016spectral,solon2019decoding}. Cross-task zero-calibration decoding (see Fig. \ref{cross-task RSVP-BCI}(a)) involves training models on EEG signals from subjects performing certain RSVP tasks (e.g., Task plane) and directly testing them on EEG data from different subjects performing distinct RSVP tasks (e.g., Task car) without using training data from the new task. Different RSVP tasks require subjects to identify varying target categories within image sequences from diverse scenes. Due to the variability in brain responses evoked by distinct image sequences across tasks \cite{wei2022erp,waytowich2016spectral}, decoding performance declines significantly when existing methods trained on data from one task are directly applied to novel RSVP tasks without cross-task calibration. This limits the rapid deployment of RSVP-BCI systems for retrieving different target categories across diverse practical scenarios. Therefore, developing an efficient cross-task zero-calibration decoding approach is essential for rapidly applying RSVP-BCI systems to novel tasks and retrieving new targets across diverse scenarios.

    To enhance cross-task zero-calibration performance, we explore the integration of additional information in the RSVP-BCI system which is related to target retrieval tasks to assist EEG decoding. In RSVP-EEG decoding, stimulus image sequences evoke EEG signals as subjects search for targets, and the decoding model detects P300 components in the signals to identify corresponding target images (see Fig. \ref{cross-task RSVP-BCI}(b)). Besides EEG signals, the stimulus images in the RSVP-BCI system inherently carry information relevant to target retrieval, which can be leveraged without additional cost since they are inherently part of the system \cite{zhang2023uav}. Thus, we propose introducing information from visual stimulus images to enhance cross-task zero-calibration decoding. However, two primary challenges remain: (1) in cross-task zero-calibration, no new stimulus images are available for training in advance for novel RSVP tasks, making it inherently a zero-shot task at the image level; and (2) typical image classification models \cite{han2022survey} primarily capture categorical information of images that cannot align with the EEG features distinguishing task-specific targets from nontargets. Recently, language-image pre-trained models have shown exceptional performance in zero-shot tasks \cite{gan2022vision}, offering a promising solution to these challenges. By leveraging language-image pre-trained models, we can utilize task-specific prompts to guide target category identification (e.g., "plane") in unseen RSVP tasks and extract task-related language-image features from both task-specific prompts and stimulus images (see Fig. \ref{cross-task RSVP-BCI}(b)), thereby complementing cross-task zero-calibration EEG decoding.
    
    In this work, we build an open-source dataset with three RSVP tasks and propose the EEG with language-image prior fusion Transformer (ELIPformer) to enhance cross-task zero-calibration decoding. First, previous research on RSVP focuses on single-task decoding, and no public RSVP-EEG dataset includes EEG signals with their corresponding stimulus images from multiple RSVP tasks. To address this, we design three distinct RSVP tasks and collect EEG data from 71 subjects. Second, considering that Transformer \cite{vaswani2017attention} can effectively capture global information on EEG signals and outperform CNN-based methods \cite{li2025temporal,song2022eeg,xie2022transformer}, we develop a Transformer-based architecture that uses metric learning to extract common RSVP-related features across different subjects. Third, we propose a prompt encoder based on the Contrastive Language-Image Pre-training (CLIP) model \cite{radford2021learning} to extract language-image features as prior knowledge to supplement and enhance EEG decoding. Finally, we conceptualize the cross-attention mechanism \cite{tsai2019multimodal} as a clustering process and propose the cross bidirectional attention (bi-attention) module to achieve efficient interaction between EEG and language-image features.
    
    In summary, the main contributions of our work can be summarized as follows:
    \begin{itemize}
        \item[1)] We design and conduct three distinct RSVP tasks and collect EEG data from 71 subjects including the corresponding stimulus image sequences. The dataset (DOI:10.57760/sciencedb.14812) is publicly available at https://doi.org/10.57760/sciencedb.14812.
    
        \item[2)] We propose the EEG with language-image prior fusion Transformer (ELIPformer) for RSVP cross-task zero-calibration decoding.  To the best of our knowledge, this is the first model that fuses EEG and language-image features in the RSVP decoding.
        
        \item[3)] We propose a prompt encoder to extract language-image features from task-specific prompts and stimulus images as prior knowledge, along with a bidirectional attention mechanism to enable effective fusion between EEG and language-image features.
        
        \item[4)] We conduct extensive experiments on the open-source RSVP cross-task dataset, which demonstrates the excellent performance of our model in the cross-task zero-calibration RSVP decoding. The source codes have been submitted as supplementary materials and will be released upon acceptance. 
    \end{itemize}
    
\section{Related Work}
    In this section, we first review the related work in RSVP decoding methods including methods proposed for subject-dependent decoding and zero-calibration decoding. Then, considering the utilization of language-image pre-training models, we provide a review of language-image pre-training methods.

\subsection{RSVP Decoding Methods}
\subsubsection{Subject-dependent Decoding} 
    The subject-dependent decoding is also known as calibration decoding or within-subject decoding, which needs to train the individual classifier with labeled data collected from the calibration procedure for each subject. Gerson et al. (2006) proposed the Hierarchical Discriminant Component Analysis (HDCA) \cite{HDCA} method to learn spatial and temporal filters that are discriminative for EEG signals. HDCA has been widely adopted in RSVP decoding due to its simplicity and efficiency. In 2010, Barachant et al. proposed the Minimum Distance to Riemannian Mean (MDRM) \cite{MDRM} for P300 decoding, which utilizes the Riemannian geometry and classifies EEG signals by using covariance matrices in Riemannian space. With the significant improvements in various fields brought about by deep learning, researchers apply deep learning methods to enhance RSVP decoding performance. Manor and Geva (2015) first proposed a CNN-based method (MCNN) \cite{MCNN} for an RSVP-based object classification task, which comprises three convolutional layers, two max-pooling layers, and three fully-connected layers. In 2018, Lawhern et al. introduced EEGNet \cite{EEGNet} for multiple EEG paradigm decoding, which applies depthwise convolution to learn spatial features and separable convolution to learn temporal features. Additionally, Zang et al. (2021) developed PLNet \cite{PLNet}, a novel CNN model that leverages the phase-locked characteristic to extract spatiotemporal features for RSVP-EEG classification. Li et al. proposed the Phase Preservation Neural Network (PPNN) \cite{PPNN}, which is capable of learning phase information and outperforms other subject-dependent methods in the P300 detection. In 2024, Ji et al. proposed a hybrid decoding model called HCANN \cite{ji2024novel}, which utilized depthwise separable convolutions to decouple the temporal dependencies between EEG signals and the multi-head self-attention mechanism to capture spatial activation patterns for RSVP decoding. 

\subsubsection{Zero-calibration Decoding}
    The zero-calibration decoding is also known as zero-training decoding or calibration-free decoding, which enables the direct use of existing classifiers on new subjects without calibration. In 2016, Waytowich et al. introduced Spectral Transfer using Information Geometry (STIG) \cite{waytowich2016spectral}, which utilizes spectral-meta learning to integrate classifiers trained on existing subjects. This method outperforms conventional subject-dependent methods when limited data is available. Lee et al. (2020) constructed a zero-calibration P300 BCI speller based on EEGNet. This approach is trained with EEG data from 55 subjects and exhibits no statistical difference in performance compared to the methods in subject-dependent decoding of online experiments \cite{LeeCNN}. Recently, Transformers have been employed to classify physiological signals. In 2022, Wang et al. proposed the Hierarchical Spatial Learning Transformer (HSLT) \cite{wang} which utilizes a Transformer-based architecture to extract discriminative spatial information from EEG data at the brain-region level for the task of emotion recognition. Furthermore, a network based on a combination of temporal convolutional network and Transformer was proposed for the classification of the sleep stage as awake or asleep \cite{TCN-T}. In the same year, Li et al. introduced the Temporal-Frequency Fusion Transformer (TFF-Former) \cite{li2022tff} for RSVP zero-calibration decoding, which separately processes the temporal and frequency information of EEG signals, and achieves significantly higher decoding performance than other zero-calibration methods. Additionally, Song et al. (2023) proposed a compact convolutional Transformer named EEGconformer, which is capable of unifying local and global features within a single EEG classification framework \cite{song2022eeg}. These studies have shown great promise of a Transformer-based network in decoding EEG signals. 

\subsection{Language-Image Pre-training Methods}
    The core idea of pre-training is to extract implicit general knowledge from a massive amount of data and subsequently transfer this knowledge to various downstream tasks \cite{han2021pre}. The language-image pre-training methods train vision models with free-form language supervision, which utilize a dual-stream architecture to learn visual and textual representations from a large number of image-text pairs and exhibit superior zero-shot ability \cite{yao2021filip,li2022grounded}. CLIP \cite{radford2021learning} has garnered unprecedented attention due to its exceptional performance in zero-shot tasks and its capacity to serve as a foundation model for vision. Many approaches inspired by CLIP have also been proposed in recent years, including Self-supervision meets Language-Image Pre-training (SLIP) \cite{mu2022slip}, which combines language supervision and self-supervision in a multi-task framework, and Fine-grained Interactive Language-Image Pre-training (FILIP) \cite{yao2021filip}, which enhances fine-grained interactions between image patches and textual words through modified contrastive loss. Additionally, Data Efficient Contrastive Language-Image Pre-training (DECLIP) \cite{li2022supervision} leverages widespread supervision in image-text pairs to learn visual features more efficiently. Building on CLIP's strengths in bridging visual and textual modalities and its exceptional generalization, our work uses CLIP to extract language-image features from task-specific prompts and stimulus images to supplement EEG decoding.
        
\section{Materials}
    We design and implement three RSVP target image retrieval tasks to collect EEG data and build the ``NeuBCI Target Retrieval RSVP-EEG Dataset”. The targets of the three tasks are plane, car, and people, respectively. For convenience, the three tasks are referred to as Task plane, Task car, and Task people, respectively. The ``NeuBCI Target Retrieval RSVP-EEG Dataset” is available at https://doi.org/10.57760/sciencedb.14812.

\subsection{Subjects}
    In the experiment, the Task plane involves 20 subjects (15 males and 5 females, aged 23.4 ± 1.1, 18 of whom are right-handed); the Task car involves 20 subjects (11 males and 9 females, aged 23.75 ± 1.3, all of whom are right-handed); and the Task people involves 31 subjects (19 males and 12 females, aged 24.9 ± 2.8, 28 of whom are right-handed). There is no overlap in subject participation across the three tasks. All participants have normal or corrected-to-normal vision and no prior experience with RSVP-based BCI experiments. Additionally, none of the participants have a history of visual disorders, neurological disease, or injury. The experimental procedures are approved by the Institutional Review Board of the Institute of Automation, Chinese Academy of Sciences, and all subjects provide written informed consent before the experiment.

\subsection{RSVP Paradigm}
\begin{figure}[!htbp] 
        \centering
        \subfigure[]{
    	\includegraphics[width=0.95\linewidth]{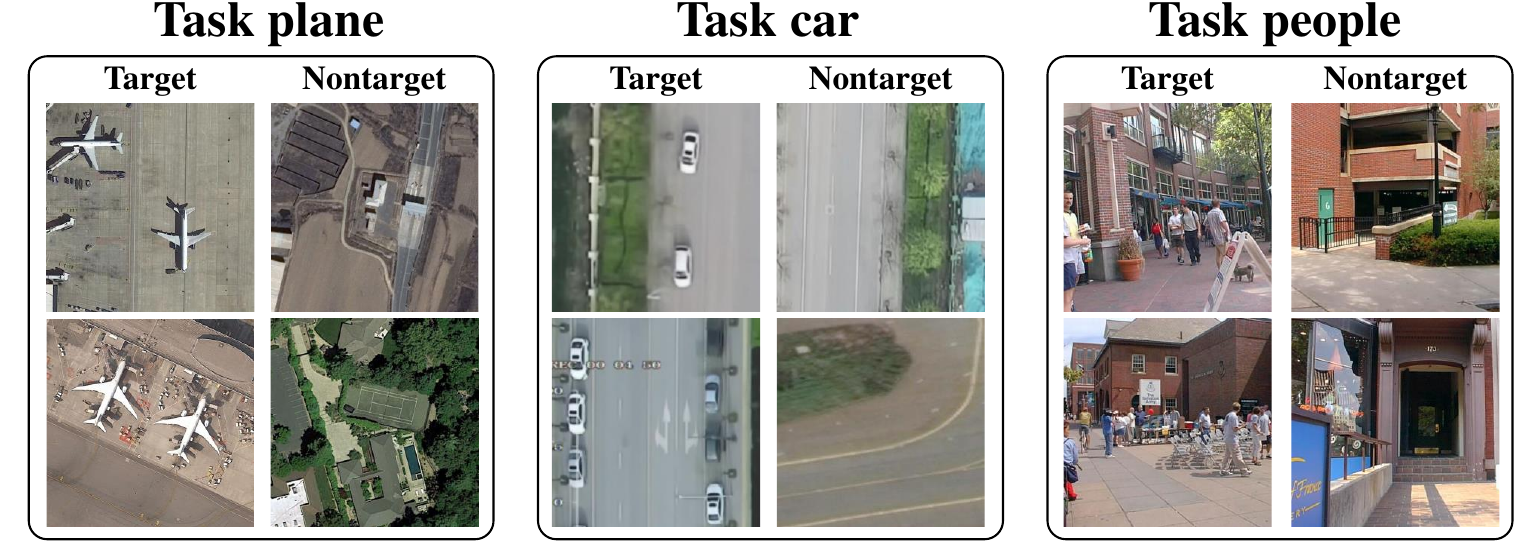}
    	}
        \subfigure[]{
    	\includegraphics[width=0.95\linewidth]{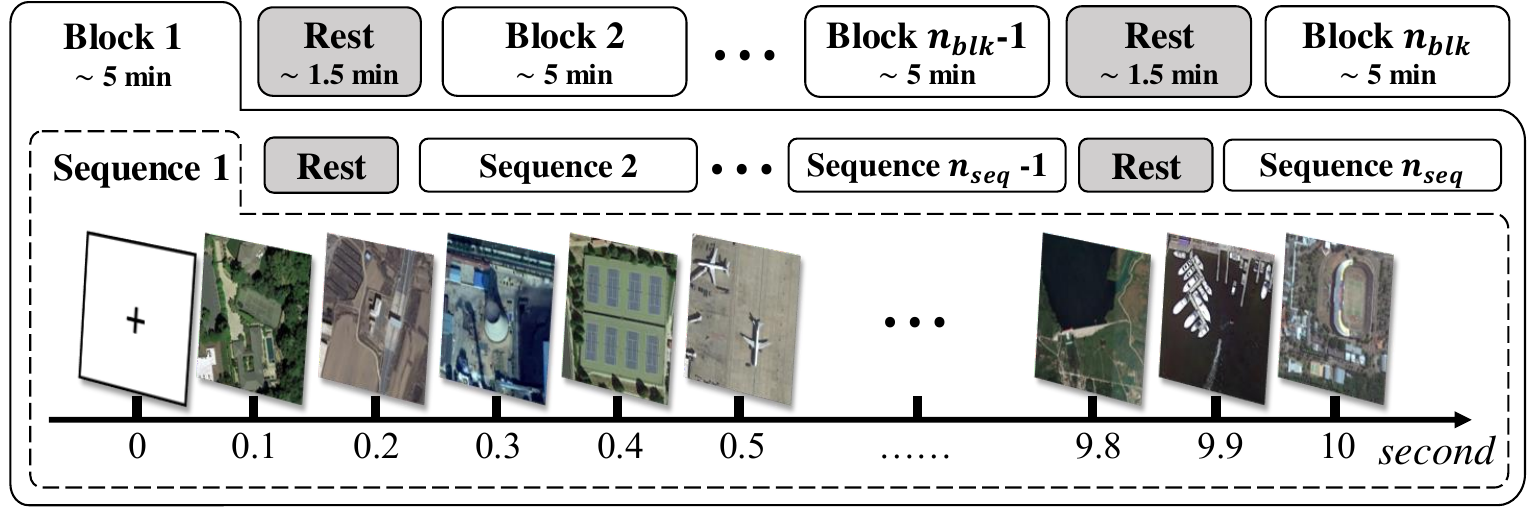}
    	}
        \caption{The RSVP-based target image retrieval experiment. (a) Examples of target and nontarget images in the three tasks, (b) settings of our RSVP experiment. The rest time between adjacent blocks is around 1-3 minutes. The rest time between adjacent sequences is controlled by subjects, around 4-6 s.}
        \label{RSVP paradigm}
    \end{figure}

    The target image retrieval experiment comprises three distinct tasks: Task plane, which requires participants to identify the image containing the target plane from remote sensing images captured by the satellite; Task car, which entails identifying the image containing the target car from ground images captured by the drone; and Task people, which requires identifying the image containing the target people from street scene images. The visual stimulus images utilized in Task plane, Task car, and Task people are from three different sources: the Dior dataset \cite{li2020object}, our self-collection drone aerial images, and the scenes and objects database \cite{torralba2009csail}, respectively. Within each task, the visual stimulus images including task-specific target are designated as target images, and the others are considered as nontarget images. Figure \ref{RSVP paradigm}(a) shows a few examples of the stimulus images.

    In the experiment, participants are seated at a distance of 1 meter from a monitor with a resolution of 1920$\times$1080. When the participants achieve a state of calmness, a sequence of images is presented randomly to them at a rate of 10Hz, and they are instructed to identify the target images. As illustrated in Fig. \ref{RSVP paradigm}(b), each subject completes a task comprising $n_{blk}$ blocks (Task plane: $n_{blk}=10$, Task car: $n_{blk}=5$, Task people: $n_{blk}=10$), with each block containing $n_{seq}$ sequences (Task plane: $n_{seq}=14$, Task car: $n_{seq}=16$, Task people: $n_{seq}=14$). Each sequence is composed of 100 images, with target images appearing at a probability of about 4\%, while the rest are non-target images. Thus, each participant observes around $n_{blk}\times n_{seq} \times 4$ target images during the experiment. The inter-sequence interval is controlled by the subjects, while each block requires around 5 minutes to complete, and the break time between adjacent blocks is approximately 2-3 minutes.

\subsection{Data Acquisition and Preprocessing}
    Data acquisition and preprocessing procedures are identical across all three tasks. The EEG data are recorded using a SynAmp2 system (NeuroScan, Australia) with 64-channel Ag$/$AgCl electrodes, placed in accordance with the International 10$/$20 system. The sampling rate is set to 1000 Hz, and the electrodes are referenced to the vertex and grounded to the AFz. The impedance of each electrode is kept below 10 $k\Omega$.

    In the preprocessing stage, the EEG data for each block are down-sampled to 250 Hz. Subsequently, the signals are filtered using a 3-order Butterworth filter with linear phase implementation between 0.1 and 15 Hz, eliminating slow drift and high-frequency noise while preventing delay distortions. The preprocessed data of each block are then segmented into EEG trials, each consisting of 1-second EEG data beginning from the onset of their stimulation to 1000 milliseconds after stimulation onset. For each EEG trial, data are channel-wise normalized to zero mean and variance one. The subsequent EEG analysis and classification are conducted based on these segmented EEG trials (samples). Therefore, each subject has a total of $n_{blk} \times n_{seq} \times 100$ EEG samples, with around $4\%$ of the samples being target samples and the remainder being nontarget samples. In the Task plane and people, each subject contains approximately 560 target samples and 13440 nontarget samples. In the Task car, each subject contains approximately 320 target samples and 7680 nontarget samples.

    \begin{figure*}[!htbp]
        \centering
        \subfigure{
            \centering
            \includegraphics[width=0.92\linewidth]{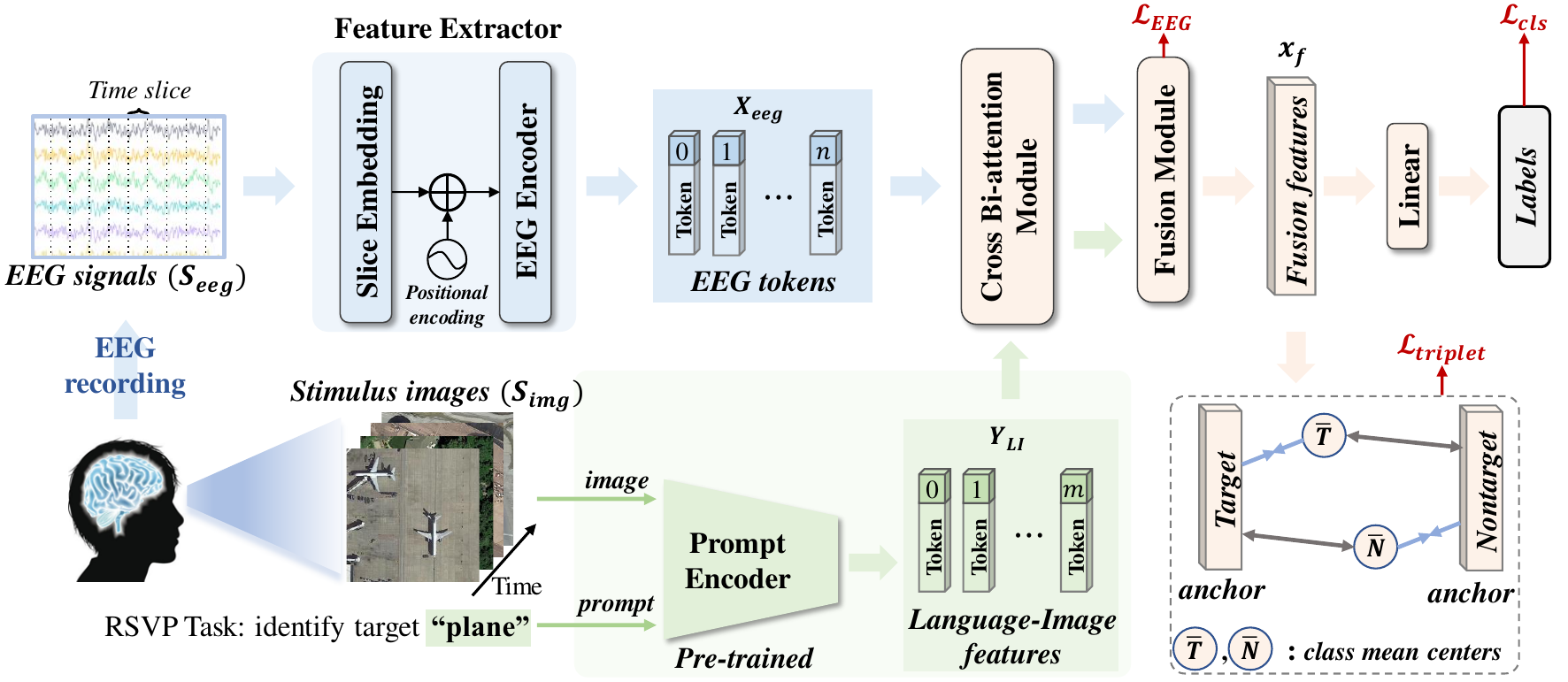}
        }
        \caption{The structure of the proposed ELIPformer. (a) ELIPformer consists of the feature extractor, the prompt encoder, the cross bi-attention module, and the fusion module, where the model takes raw EEG signals ($\boldsymbol{S}_{eeg}$), corresponding stimulus images ($\boldsymbol{S}_{img}$), and task-specific prompts as input. Initially, the feature extractor extracts EEG features ($\boldsymbol{X}_{eeg}$), and the prompt encoder extracts language-image features ($\boldsymbol{Y}_{LI}$), respectively. Subsequently, the cross bi-attention module facilitates the modal interaction between extracted EEG and image tokens. Finally, the fusion module combines the output EEG and image tokens as fusion features ($\boldsymbol{x}_{f}$) for classification.}
        \label{model framework}
    \end{figure*}
  
\section{Method}
   The proposed EEG with language-image prior fusion Transformer (ELIPformer) consists of a feature extractor, a prompt encoder, a cross bi-attention module, and a fusion module (see Fig. \ref{model framework}). The model takes raw EEG signals ($\boldsymbol{S}_{eeg}\in \mathbb{R}^{C\times T}$), corresponding stimulus images ($\boldsymbol{S}_{img}\in \mathbb{R}^{H\times W \times 3}$), and task-specific prompts as input, where the prompts specify the target class to be identified in the current RSVP task. The $C$ denotes the number of channels, $T$ denotes the number of time points, and $H$ and $W$ denote the height and width of images, respectively. The feature extractor extracts EEG temporal features through slice embedding and realizes global interaction among temporal dimensions. The prompt encoder extracts language-image features that distinguish targets and nontargets in the current RSVP task as prior knowledge. Subsequently, the cross bi-attention module facilitates feature interaction between EEG features and language-image features to bridge the semantic gap between them. Finally, the fusion module integrates features from both modalities to produce more informative and discriminative representations for classification.
        
\subsection{Feature Extractor}
    We design a feature extractor consisting of three parts: a slice embedding layer, a position embedding layer, and an encoder layer. Since the characteristics of RSVP-EEG are mainly reflected in the temporal dimension, we partition and transform the input raw EEG signals ($\boldsymbol{S}_{eeg}$) into a sequence of flattened 2-dimensional slices ($\boldsymbol{s}_{i}\in \mathbb{R}^{(C\times t)}, i = 1,2,\cdots,n_{s}$) each representing the brain activity within a time period, where $C$ denotes the number of channels, and $T$ denotes the number of time points, $t$ denotes the slice length, and $n_{s} = [T /t]$ is the number of slices which also serves as the input length for the encoder layers. Then we flatten the slices and map them to $d_{model}$ dimensions with a trainable linear projection, where $d_{model}$ is a constant latent vector size across all layers in our model:
    \begin{equation}
        \boldsymbol{X}_{slice} = [\boldsymbol{s}_{1};\boldsymbol{s}_{2},\cdots;\boldsymbol{s}_{n_{s}}]^{T}\boldsymbol{W}+ \boldsymbol{W}_{pos},
    \end{equation}
    where $\boldsymbol{W}\in \mathbb{R}^{(C\times t)\times d_{model}}$ denotes the linear projection matrix and $\boldsymbol{W}_{pos}\in \mathbb{R}^{n_{s}\times d_{model}}$ denotes the learnable positional parameters which used to preserve the temporal information. The output of the slice and position embedding layer ($\boldsymbol{X}_{slice}\in \mathbb{R}^{n_{s}\times d_{model}}$) are referred to as EEG tokens which are inputs to the EEG encoder layer.
 
    The EEG encoder layer includes two sub-layers: a Multi-head Self-Attention (MSA) layer that captures global relationships among input tokens, and a position-wise Feed-Forward Network (FFN) layer that extracts feature representation using a Multi-Layer Perceptron (MLP) with one hidden layer of $h\times d_{model}$ hidden dimensions. All sub-layers utilize a residual connection and a layer normalization operation to enhance the scalability of the Transformer. The GELU \cite{hendrycks2016gaussian} is used as the activation function throughout the model. Moreover, we add a skip connection between the start and end of the encoder layer. The EEG encoder layer outputs $\boldsymbol{X}_{eeg}\in \mathbb{R}^{n_{s}\times d_{model}}$, which has the same dimensions as the input $\boldsymbol{X}_{slice}$. It captures comprehensive temporal information from the EEG signals via the global feature interaction achieved by the self-attention mechanism.

\subsection{Prompt Encoder}
    To enhance the decoding performance of the RSVP-BCI system in cross-task zero-calibration, we propose to extract language-image features from task-specific prompts and stimulus images using the pre-trained model as prior knowledge to supplement EEG decoding. Initially, we intend to utilize the pre-trained vision models \cite{dosovitskiy2020image,liu2021swin} to extract features from stimulus images as prior knowledge. However, the pre-trained vision models primarily capture image category information, while the EEG features extracted by EEG decoding models focus on distinguishing task-specific targets from nontargets. This discrepancy leads to a semantic mismatch between the EEG features and the extracted image features. To address this issue, we propose a prompt encoder based on CLIP that informs the pre-trained vision model about the target category (i.e., plane, car, or people) for the current RSVP task by providing task-specific prompts to the model.

    \begin{figure}[!htbp]
        \centering
        \centering
        \includegraphics[width=0.98\linewidth]{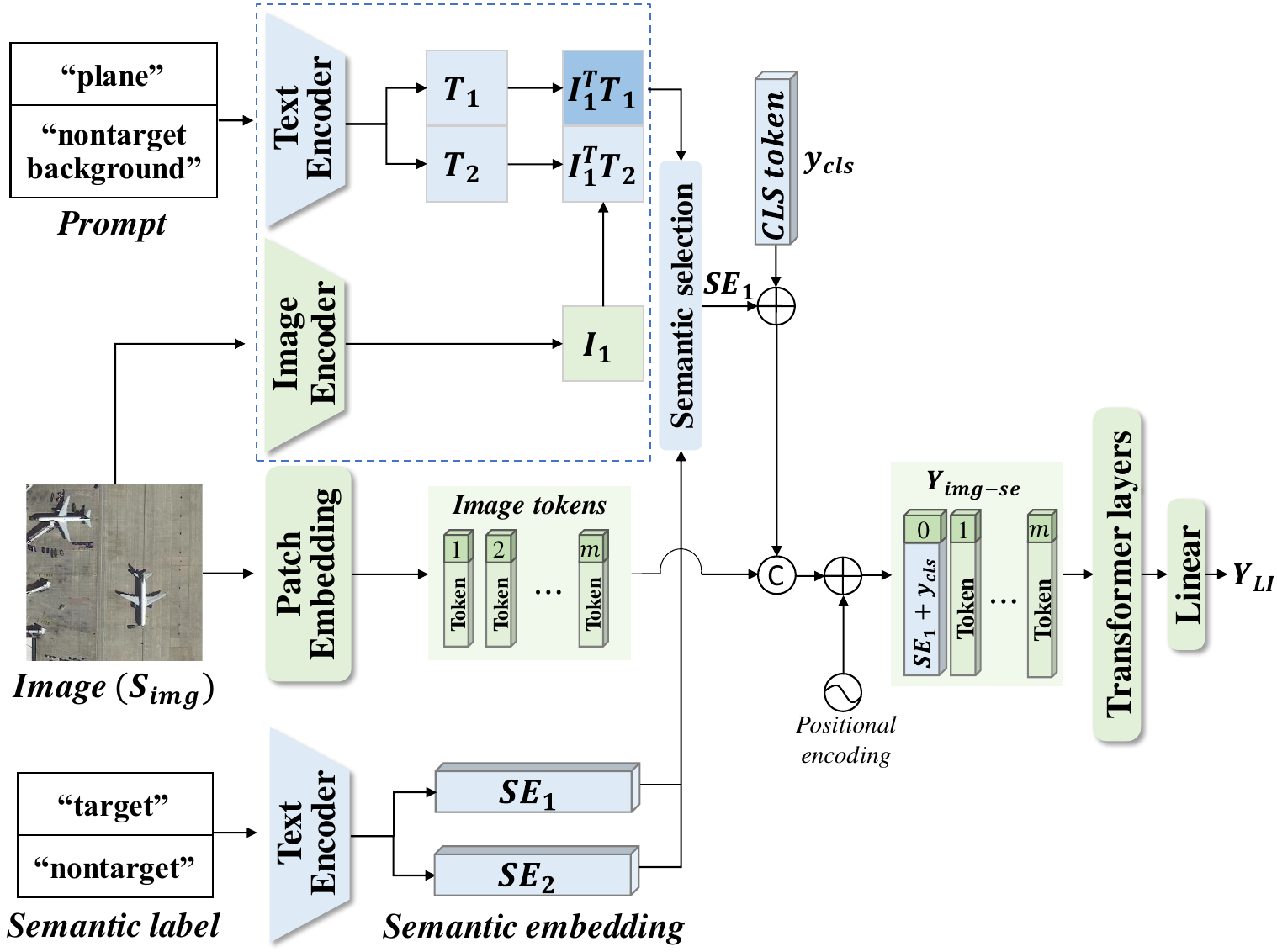}
        \caption{The structure of the prompt encoder. The prompt encoder consists of components from the pre-trained CLIP-ViT-B/32 model \cite{radford2021learning}. Both the image encoder and text encoder are inherited from this model. Additionally, the patch embedding layer and transformer layers are derived from the image encoder in CLIP-ViT-B/32.}
        \label{prompt encoder}
    \end{figure}

    As depicted in Fig. \ref{prompt encoder}, the prompt encoder takes the image ($\boldsymbol{S}_{img}$) and the prompt specifying the target category in the current RSVP task as inputs. Since the cross-task RSVP decoding is a binary classification task and the target category is the only information available before performing a new task, the input prompt includes two classes: ``[the target category]" for the target and the fixed ``nontarget background" for the nontarget. At first, the pre-trained image encoder and text encoder of CLIP are utilized to extract the features from the input image ($\boldsymbol{I}_{1}\in \mathbb{R}^{d_{enc}}$) and prompt ($\boldsymbol{T}_{i}\in \mathbb{R}^{d_{enc}}, i=1, 2$), where $d_{enc}=512$ is the output dimension of the image and text encoders. Subsequently, the cosine similarities $sim\left(\boldsymbol{I}_{1},\boldsymbol{T}_{i}\right) = \boldsymbol{I}_{1}^{T}\boldsymbol{T}_{i} / ||\boldsymbol{I}_{1}||_{2} ||\boldsymbol{T}_{i}||_{2}, i=1,2)$ between image and prompt features are calculated. The image encoder utilizes the Vision Transformer (ViT) \cite{dosovitskiy2020image} with a patch size of 32. Based on these similarities, the corresponding semantic embedding is selected to distinguish between target and nontarget, where the semantic embedding ($\boldsymbol{SE}_{i}\in \mathbb{R}^{d_{enc}}, i=1, 2$) of the target class and nontarget class is the features of ``target" and ``nontarget" extracted by the text encoder, respectively. Then, the class token and the selected semantic embedding are summed as the new class token of the image: 
    \begin{equation}
        \boldsymbol{y}_{cls-se} = \boldsymbol{y}_{cls} + \sum_{i=1}^{2} \sum_{j=1, j \neq i}^{2} \mathbb{I}_{\left( sim \left(\boldsymbol{I}_{1}, \boldsymbol{T}_{i}\right) > sim \left(\boldsymbol{I}_{1}, \boldsymbol{T}_{j}\right) \right)} \boldsymbol{SE}_{i},
    \end{equation}
    where $\boldsymbol{y}_{cls}$ represents the pre-trained image class token in ViT, and the $\mathbb{I}_{(\cdot)}$ is an indicator function that outputs 0 or 1 based on the subscript condition. The semantically embedded class token $\boldsymbol{y}_{cls-se}$ is then used as the new class token for patch embedding the input image ($\boldsymbol{S}_{img}$) within the patch layer of CLIP to generate image tokens with semantic embedding. The new class token $\boldsymbol{y}_{cls-se}$ combined with semantic embedding ($\boldsymbol{SE}_{i}, i=1, 2$) aggregates the information that distinguishes between target and nontarget in the subsequent Transformer layers. The process of generating image tokens with semantic embedding ($\boldsymbol{Y}_{img-se}$) can be formulated as follows:
    \begin{equation}
        \boldsymbol{Y}_{img-se} = Concat\left[ \boldsymbol{y}_{cls-se}, p(\boldsymbol{S}_{img}) \right] + \boldsymbol{W}_{pos-clip},
    \end{equation}
    where $p(\cdot)$ is the image patch embedding operation in the image encoder of CLIP, and $\boldsymbol{W}_{pos-clip}$ represents the positional parameters in the image encoder of CLIP. Finally, the image tokens with semantic embedding ($\boldsymbol{Y}_{img-se}$) are fed into the pre-trained Transformer layers in the image encoder of CLIP, followed by a linear layer to project the dimension of the tokens to $d_{model}$. The outputs of the prompt encoder are language-image features ($\boldsymbol{Y}_{LI} \in \mathbb{R}^{n_{s}\times d_{model}}$) extracted from task-specific prompts and stimulus images. 
    
\subsection{Cross Bi-attention Module}
\begin{figure}[!htbp]
        \centering
        \centering
        \includegraphics[width=0.77\linewidth]{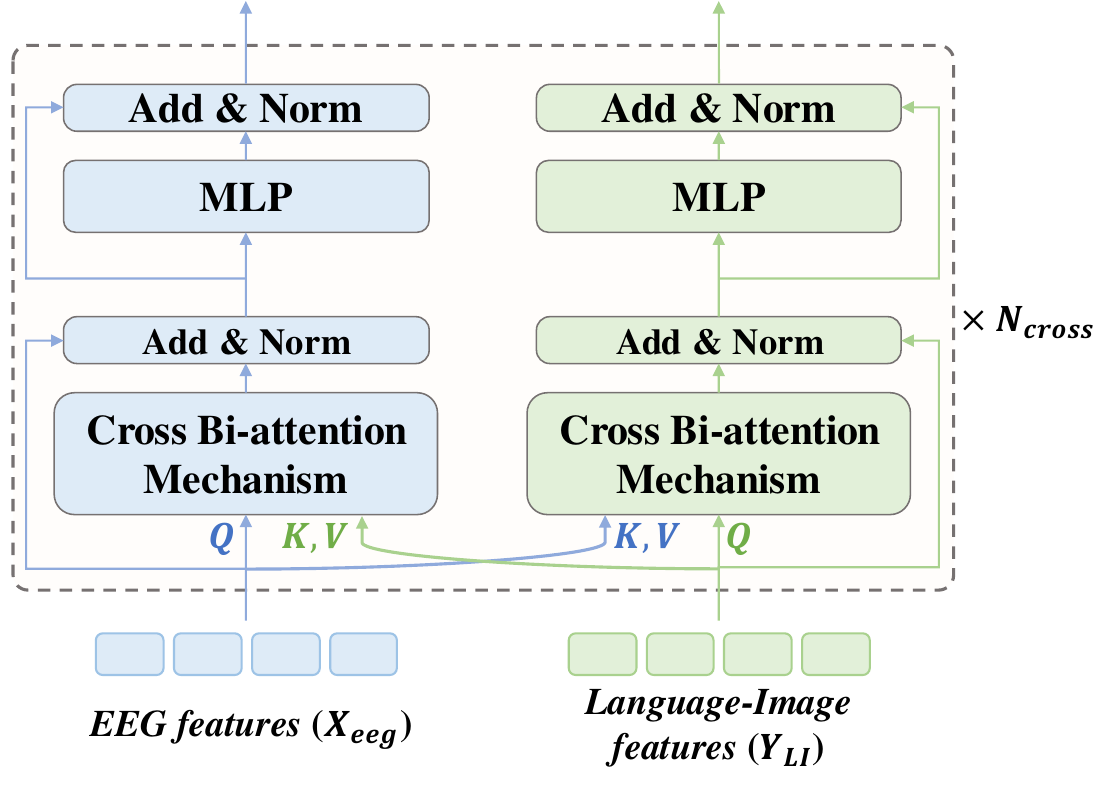}
        \caption{The structure of cross bi-attention module. The cross bi-attention module is composed of $N_{cross}$ successive cross bi-attention layers for effective interaction between EEG features ($\boldsymbol{X}_{eeg}$) and language-image features ($\boldsymbol{Y}_{LI}$).}
        \label{cross-attention}
    \end{figure}

    To enable efficient feature alignment and interaction between EEG and language-image features, we introduce a cross bidirectional attention module. Traditional cross-attention mechanism \cite{tsai2019multimodal} allows one modality to query another to compute attention weights for feature interaction, but this approach may reduce interaction efficiency \cite{ramachandran2019stand}. To enhance the efficiency of modal interaction, we propose a cross bidirectional attention mechanism that reformulates the attention process as a Gaussian mixture clustering task, calculating attention weights from both query modality and key modality perspectives to improve cross-modal interaction. The cross bi-attention module consists of $N_{cross}$ successive layers, which replace the MSA layer in the EEG encoder with a multi-head cross bidirectional attention (MHCBA) layer. At the same time, the rest of the structure remains unchanged (see Fig. \ref{cross-attention}).

\subsubsection{Analysis of Cross-attention Mechanism}
    In the cross-attention mechanism, linear projection maps input tokens of two inputs $\boldsymbol{X}_{eeg}\in \mathbb{R}^{n_{s}\times d_{model}}$ and $\boldsymbol{Y}_{LI}\in \mathbb{R}^{n_{s}\times d_{model}}$ to three different sequential vectors (query $\boldsymbol{Q}_{x}\in \mathbb{R}^{n_{s}\times d_{k}}$, key $\boldsymbol{K}_{y}\in \mathbb{R}^{n_{s}\times d_{k}}$, and value $\boldsymbol{V}_{y}\in \mathbb{R}^{n_{s}\times d_{v}}$), respectively. The interaction of $\boldsymbol{X}_{eeg}$ guiding $\boldsymbol{Y}_{LI}$ is formulated as:
    \begin{equation}
        \begin{split}
            \hat{\boldsymbol{X}}_{eeg} &= \boldsymbol{Q}_{x} + \sigma(\boldsymbol{Q}_{x}\boldsymbol{K}_{y}^{T})\boldsymbol{V}_{y}  \\
            & = \boldsymbol{X}_{eeg}\boldsymbol{W}_{Q} + \sigma(\boldsymbol{X}_{eeg}\boldsymbol{W}_{qk}\boldsymbol{Y}_{LI}^{T})\boldsymbol{Y}_{LI}\boldsymbol{W}_{V},
        \end{split}
        \label{cross-attention equation}
    \end{equation}
    where the $\sigma(\cdot)$ denotes the softmax function. The $\boldsymbol{W}_{qk} = \boldsymbol{W}_{Q}\boldsymbol{W}_{K}^{T}$ where the $\boldsymbol{W}_{Q}, \boldsymbol{W}_{K}\in \mathbb{R}^{d_{model}\times d_{k}},\boldsymbol{W}_{V}\in \mathbb{R}^{d_{model}\times d_{v}}$ are learnable parameters. For simplicity, the scaling factor $1/\sqrt{d_{k}}$ is omitted. In the cross bi-attention module, EEG and language-image features mutually guide interaction with each other according to the similarity between their tokens.

    In Eq. (\ref{cross-attention equation}), the process of cross-attention is similar to the clustering, where tokens ($\boldsymbol{x}_{i}\in\mathbb{R}^{d_{model}}$) in $\boldsymbol{X}_{eeg} = [\boldsymbol{x}_{1}, \boldsymbol{x}_{2}, \cdots, \boldsymbol{x}_{n_{s}}]^{T}$ serve as the cluster centers and tokens ($\boldsymbol{y}_{j}\in\mathbb{R}^{d_{model}}$) in $\boldsymbol{Y}_{LI} = [\boldsymbol{y}_{1}, \boldsymbol{y}_{2}, \cdots, \boldsymbol{y}_{n_{s}}]^{T}$ as the cluster samples. The $i$-th row of the $\hat{\boldsymbol{X}}_{eeg} = [\hat{\boldsymbol{x}}_{1}, \hat{\boldsymbol{x}}_{2},\cdots,\hat{\boldsymbol{x}}_{n_{s}}]^{T}$ can be formulated as:
    \begin{equation}
        \begin{split}
            \hat{\boldsymbol{x}}_{i}^{T} = \boldsymbol{x}_{i}^{T}\boldsymbol{W}_{Q} +\sum_{j=1}^{n_{s}}\dfrac{\exp({\boldsymbol{x}_{i}^{T} \boldsymbol{W}_{qk} \boldsymbol{y}_{j}})}{\sum_{t=1}^{n_{s}}\exp({\boldsymbol{x}_{i}^{T} \boldsymbol{W}_{qk} \boldsymbol{y}_{t}})}\boldsymbol{y}_{j}^{T}\boldsymbol{W}_{V}.
        \end{split}
    \end{equation}
    This process can be regarded as an update of the clustering center in the clustering problem, where the token $\boldsymbol{x}_{i}$ acts as the cluster center, and tokens $\boldsymbol{y}_{j}$ serve as cluster samples. The similarity score between $\boldsymbol{x}_{i}$ and each $\boldsymbol{y}_{j}$ determines the contribution of $\boldsymbol{y}_{j}$ to the updated cluster center $\hat{\boldsymbol{x}}_{i}$. By aggregating the weighted features of the samples $\{\boldsymbol{y}_{j}\}_{j=1}^{n_s}$, the cluster center $\boldsymbol{x}_{i}$ is refined to better represent its surrounding cluster samples, particularly those with higher similarity to $\boldsymbol{x}_{i}$.

\subsubsection{Cross Bidirectional Attention Mechanism}
    To enhance the cross-attention mechanism, we employ the clustering center updating algorithm of the classical Gaussian Mixture Model \cite{mclachlan1988mixture}, which is derived from the Expectation-Maximization (EM) algorithm \cite{dempster1977maximum} and consists of two steps. First, the posterior probability that a given sample belongs to each cluster center is computed. Subsequently, the normalized posterior probabilities are used as weights to perform a weighted summation of samples. In the cross-attention mechanism, the softmax similarity between a sample $\boldsymbol{y}_{j}$ and each cluster center $\boldsymbol{x}_{i}$ is utilized to approximate the posterior probability. These similarities are subsequently normalized and used to update tokens $\boldsymbol{x}_{i}$ through weighted summation. For example, the update for the $i$-th token in $\boldsymbol{X}_{eeg}$ is formulated as follows:
    \begin{equation}
        \hat{\boldsymbol{x}}_{i}'^{T} =\boldsymbol{x}_{i}^{T}\boldsymbol{W}_{Q} + \frac{1}{N_{i} }\sum_{j=1}^{n_{s}}\dfrac{\exp({\boldsymbol{y}_{j}^{T} \boldsymbol{W}_{qk}^{T} \boldsymbol{x}_{i}})}{\sum_{t=1}^{n_{s}}\exp({\boldsymbol{y}_{j}^{T} \boldsymbol{W}_{qk}^{T} \boldsymbol{x}_{t}})}   \boldsymbol{y}_{j}^{T}\boldsymbol{W}_{V},
        \label{bi-attention equation}
    \end{equation}
    \begin{equation}
        N_{i} = \sum_{r=1}^{n_{s}}\dfrac{\exp({\boldsymbol{y}_{r}^{T} \boldsymbol{W}_{qk}^{T} \boldsymbol{x}_{i}})}{\sum_{t=1}^{n_{s}}\exp({\boldsymbol{y}_{r}^{T} \boldsymbol{W}_{qk}^{T} \boldsymbol{x}_{t}})}.
    \end{equation}

    Finally, the interaction of EEG features guiding language-image features (illustrated in the cross bi-attention mechanism on the left of Fig. \ref{cross-attention}) is achieved by summing the tokens generated by the two attention mechanisms in Eq. (\ref{cross-attention equation}) and Eq. (\ref{bi-attention equation}), which is ($\hat{\boldsymbol{x}}_{i}+\hat{\boldsymbol{x}}'_{i}$). In this enhanced mechanism, when the $i$-th token $\boldsymbol{x}_{i}$ in $\boldsymbol{X}_{eeg}$ is updated, the weights of the tokens of another modality $\{\boldsymbol{y}_{j}\}_{j=1}^{n_{s}}$ are computed based on two factors: (1) the similarity between $\boldsymbol{x}_{i}$ and all tokens in $\boldsymbol{Y}_{LI}$ ($\boldsymbol{x}_{i}^{T} \boldsymbol{W}_{qk} \boldsymbol{y}_{j}, j=1, \cdots, n_{s}$). (2) the similarity between $\boldsymbol{y}_{j}$ and all tokens in $\boldsymbol{X}_{eeg}$ ($\boldsymbol{y}_{j}^{T} \boldsymbol{W}_{qk}^{T} \boldsymbol{x}_{i}, i=1, \cdots, n_{s}$). The weight of each token in $\boldsymbol{Y}_{LI}$ is determined from the perspective of $\boldsymbol{X}_{eeg}$ and $\boldsymbol{Y}_{LI}$, respectively. Similarly, the interaction of language-image features guiding EEG features (illustrated in the cross bi-attention mechanism on the right of Fig. \ref{cross-attention}) follows the same process, where all roles of $\boldsymbol{X}_{eeg}$ and $\boldsymbol{Y}_{LI}$ are reversed, such that the tokens in $\boldsymbol{X}_{eeg}$ are weighted based on their similarity to tokens in $\boldsymbol{Y}_{LI}$. Therefore, we refer to this enhanced mechanism as the cross bidirectional attention mechanism, which enables the query modality to obtain more information from another modality, achieving efficient interaction between the two modalities. 
    
\subsection{Fusion Module}
    After the cross bi-attention module, we employ a fusion module including an EEG encoder layer and a convolutional layer to aggregate all EEG tokens and concatenate them with the image class token for feature fusion. The input to the fusion module comprises the EEG tokens and the image tokens output from the cross bi-attention module. Specifically, the global feature ($\boldsymbol{x}_{eeg}\in\mathbb{R}^{d_{model}}$) of the EEG tokens is extracted and flattened by an EEG encoder layer and a convolutional layer. The EEG encoder layer in the fusion module is identical to the one in the feature extractor. The convolutional layer consists of 16 convolution kernels with a kernel size of $(n_{s}, d_{model}/8)$ and the stride is equal to the kernel size, which ensures the extracted feature remains $d_{model}$ dimensions after flattening. Then the EEG global feature is concatenated with the class token of the image tokens as the fusion feature ($\boldsymbol{x}_{f}\in\mathbb{R}^{2d_{model}}$) for classification.
  
    Since the prompt encoder is pre-trained while the EEG feature extractor is trained from scratch, unbalanced optimization may occur during training. To address this, the EEG loss is introduced to balance the optimization process. The EEG loss is a cross-entropy loss and is computed after projecting $\boldsymbol{x}_{eeg}$ to two dimensions via a linear layer. The formulation for the EEG loss is as follows:
    \begin{equation}
        \mathcal{L}_{EEG} = -\dfrac{1}{N}\sum_{n=1}^{N}\sum_{r=1}^{R}y_{n,r}\log \hat{y}_{n,r}^{eeg},
    \end{equation}
    where $R$ denotes the number of classes, $y$ indicates the real label and $\hat{y}_{n,r}^{eeg}$ is the prediction result using $\boldsymbol{x}_{eeg}$. 
    
    Moreover, to improve the discrimination of fusion features across classes, we use the triplet loss \cite{balntas2016learning}. For each training batch, this loss minimizes the distance between fusion features and the mean center of their corresponding class while maximizing the distance from other class mean centers. The triplet loss is formulated as:
    \begin{equation} 
        \begin{split}
            \mathcal{L}_{triplet}& = \frac{1}{N} \sum_{i=1}^{N} \bigg[ \|\boldsymbol{x}_{f}^{i} - \frac{1}{\sum_{t=1}^{N}\delta_{y_{t}y_{i}}}\sum_{j=1}^{N}\delta_{y_{j}y_{i}}\boldsymbol{x}_{f}^{j} \|_{2}^{2} \\
            &- \|\boldsymbol{x}_{f}^{i} - \frac{1}{\sum_{t=1}^{N}\delta_{y_{t}(1-y_{i})}}\sum_{j=1}^{N}\delta_{y_{j}(1-y_{i})}\boldsymbol{x}_{f}^{j} \|_{2}^{2} + \alpha  \bigg]_{+},
        \end{split}
    \end{equation}
    where $\|\cdot\|_{2} $ is the Euclidean distance and $\left[ \cdot \right]_{+}$ denotes $\max(\cdot,0)$. $\delta_{ij}$ denotes the Kronecker delta, which equals 1 if $i=j$ and 0 otherwise.
    
    The overall loss function comprises the EEG loss, triplet loss, and classification loss. The classification loss is computed after projecting $\boldsymbol{x}_{f}$ to two dimensions using a linear layer. The classification loss can be formulated as follows:

    \begin{equation}
        \mathcal{L}_{cls} = -\dfrac{1}{N}\sum_{n=1}^{N}\sum_{r=1}^{R}y_{n,r}\log \hat{y}_{n,r}^{f},
    \end{equation}
    where $\hat{y}_{n,r}^{f}$ is the prediction result based on $\boldsymbol{x}_{f}$. Finally, the overall loss for the model training is: 
    \begin{equation}
        \mathcal{L}_{overall} = \mathcal{L}_{cls} + \mathcal{L}_{triplet} + \mathcal{L}_{EEG}.
    \end{equation} 
    
\section{Experiments}

\subsection{Hyperparameter Settings and Implementation Details}
\subsubsection{Hyperparameter Settings}
    The symbols and values of network hyperparameters in ELIPformer are summarized in Table \ref{hyperparameters}. 

    \setlength{\tabcolsep}{2.0mm}{
	\begin{table}[htbp]
            \small
            \centering
            \renewcommand\arraystretch{1.2}
            \caption{The Symbols and Values Summary of Network Hyperparameters.}
            \label{hyperparameters}
            \begin{tabular}{cp{5.7cm}l}
                \toprule[1.2pt]
                \textbf{Symbol} & \textbf{Definition}  & \textbf{Value} \\
                \hline
                $C$ & Number of EEG channels  & 64 \\
                $T$ & Number of EEG sampling points & 250 \\
                $H, W$ & Height and width of input images & 224 \\
                $t$ & Slice length in the slice embedding layer  & 5 \\
                $n_{s}$ & Length of token sequence & 50 \\
                $d_{model}$ & Embedding dimension of all the tokens & 128 \\
                $h$ & Number of attention heads in self-attention and cross bi-attention mechanisms & 4 \\
                $N_{cross}$ & The number of cross bi-attention layers & 2 \\
                \bottomrule[1.2pt] 
            \end{tabular}
	\end{table}} 

    We also set the latent feature dimensions ($d_{k}$, $d_{v}$) in the attention mechanisms equal to $d_{model}$ for both self-attention and cross bi-attention mechanisms. To prevent overfitting, we reduce the feature embedding dimension $d_{model}$ to 128 and the number of attention heads $h$ to 4 compared to the original Transformer \cite{vaswani2017attention}.

\subsubsection{Implementation Details}
    We implement the proposed ELIPformer using PyTorch \cite{paszke2019pytorch}. In both the training and testing stages, the input to the ELIPformer includes three components: the EEG signals, the corresponding stimulus images, and the task-specific prompts. The preprocessing of the EEG signals is described in Section III.C, with an input size of 64 (channels)$\times$250 (sampling rate). The stimulus images are resized to 3$\times$224$\times$224, and task-specific prompts are defined based on the target categories identified by the subjects in the RSVP task. The prompts for targets and nontarget are set as ``[the target category]" and ``nontarget background", respectively. For instance, in the Task plane, the prompts are ``plane" for targets and ``nontarget background" for nontargets.
    
    Since the prompt encoder uses pre-trained models while the EEG decoding model is trained from scratch, this creates an imbalance in the training process. To address this issue, the training process is divided into two stages. In the first stage, the EEG decoding model is pre-trained by minimizing the EEG loss $\mathcal{L}_{EEG}$. In the second stage, the entire network is optimized by minimizing the overall loss $\mathcal{L}_{overall}$, where the margin ($\alpha$) in the triplet loss $\mathcal{L}_{triplet}$ is set to 0.5. Model optimization is performed using the Adam optimizer \cite{kingma2014adam} with an initial learning rate of 0.001, reduced by 20\% every 10 epochs in the first stage and every 20 epochs in the second. We apply $L2$ regularization with a weight decay coefficient of 0.01. To ensure robustness in the triplet loss mean center, The batch size ($N$) is set to 64 in the first stage and 1024 in the second, with a maximum of 30 and 50 epochs, respectively.
    
\subsection{Comparison Methods}
    The compared EEG decoding methods are divided into three categories, including conventional methods, CNN-based methods, and Transformer-based methods.
    
    To ensure fair comparisons, we resample our EEG signals as described in the comparison methods. The comparison methods without available open-source codes are re-implemented exactly, and the models are optimized using the same techniques and parameter settings as detailed in the source literature. The comparison methods are as follows:

\subsubsection{Conventional machine learning methods}
    \begin{itemize}
        \item HDCA \cite{HDCA}: a linear discrimination method that learns weights on EEG channels and time windows. The time window is set to 25 sampling points.
        
        \item MDRM \cite{MDRM}: a Riemannian geometry classifier based on the geodesic distance to category centers. The method is re-implemented with the Python package pyriemann.
    \end{itemize}
    
\subsubsection{CNN-based methods}
    \smallskip
    \begin{itemize}
        \item EEGNet \cite{EEGNet}: a CNN-based method with depthwise and separable convolution layers. We utilize the open-source implementation in https://github.com/vlawhern/arl-eegmodels.
        
        \item LeeNet \cite{LeeCNN}: a CNN based on EEGNet, which is suitable for large ERP datasets. This method is re-implemented according to \cite{LeeCNN}.
        
        \item PLNet \cite{PLNet}: a CNN leveraging phase-locked characteristics of ERP signals to extract spatiotemporal features. This method is re-implemented according to \cite{PLNet}.
        
        \item PPNN \cite{PPNN}: a CNN with dilated temporal convolution layers to preserve phase information. This method is re-implemented according to \cite{PPNN}.
        
        \item HCANN \cite{ji2024novel}: a CNN-based method combining depthwise separable convolution with multi-scale factors and the attention mechanism to capture temporal and spatial EEG representations. We utilize the open-source implementation at https://github.com/youshuoji/HCANN.
    \end{itemize}

\subsubsection{Transformer-based methods}
    \smallskip
    \begin{itemize}
        \item HSLT \cite{wang}: a Transformer-based model that captures EEG spatial dependencies at electrode and brain-region levels. This method is re-implemented according to \cite{wang}.
        
        \item TCN-T \cite{TCN-T}: a network combining Temporal Convolution Network and Transformer. We utilize the open-source implementation at https://github.com/rcasal/sleep\_transformer.
        
        \item TFF-Former \cite{li2022tff}: a two-stream Transformer-based network that extracts EEG temporal-spatial and frequency-spatial features. The implementation refers to \cite{li2022tff}.
        
        \item EEG-conformer \cite{song2022eeg}: a compact convolutional Transformer that encapsulates local and global features in a unified framework. We utilize the open-source implementation at https://github.com/eeyhsong/EEG-Conformer.
        
        \item EEG baseline: a component of ELIPformer including a feature extractor, an encoder layer, and a convolution layer within the fusion module. The EEG baseline model uses only EEG signals and is optimized using the loss function $\mathcal{L}_{cls}$.
    \end{itemize}

\subsection{Experimental Setup}
    In Section III, we design three distinct RSVP experiments for different application scenarios. There is no overlap between subjects on the different tasks. As a result, Task plane, Task car, and Task people can be considered independent RSVP datasets. To evaluate the performance of our proposed model against existing RSVP decoding models, we perform cross-task zero-calibration experiments on these three tasks. Specifically, the models are trained on the data from all subjects in one task and then tested on each subject in the other two tasks. This setup results in six comparative cross-task experiments (Training task $\to$ Test task: car$\to$plane, people$\to$plane, plane$\to$car, people$\to$car, plane$\to$people, car$\to$people). For instance, in the car$\to$plane experiment, the model is trained on data from all subjects in the Task car and tested on each subject in the Task plane, where the average performance across all subjects in the Task plane is reported as the final result. Additionally, to address the issue of extreme class imbalance in RSVP tasks, we down-sample the nontarget class to match the size of the target class only in the training set.
    
\subsection{Evaluation Metrics and Statistical Analysis}
    For evaluation metrics, we employ the balanced-accuracy (BA), which fits the class imbalance dataset. The calculation formula for BA is as follows:
    
    \begin{equation}
	    BA = \left( \frac{TP}{TP + FN} + \frac{TN}{TN + FP} \right) /2 ,
    \end{equation}
    where the number of positive samples that are correctly classified is referred to as TP, and FN refers to the number of positive samples that are incorrectly classified. TN denotes the number of negative samples that are correctly classified, while FP represents the number of negative samples that are incorrectly classified. The EEG decoding results are presented as mean $\pm$ standard deviation across all test subjects. 
 
    Moreover, we employ the one-way analysis of variance (ANOVA), two-way ANOVA, and two-way repeated measures ANOVA to assess the influence of different factors on RSVP decoding performance. The Greenhouse-Geisser correlation is applied in repeated measures ANOVA if the data do not conform to the Sphericity assumption by Mauchly’s Test of Sphericity. Post-hoc analysis for each significant factor is performed and the Holm-Bonferroni correction is applied to adjust the p-value in all post-hoc pairwise comparisons. In all statistical analyses, the significance level is set at 0.05. 

\section{Results and Discussion}
\subsection{Comparative Experiment}
    To compare the performance of our proposed model and existing RSVP decoding models, we conduct six cross-task zero-calibration experiments on our dataset consisting of three tasks. The classification results for different comparison methods in six experiments are presented in Table \ref{comparison experiments cross-task zero-calibration}. For instance, car$\to$plane denotes that the model is trained on data from Task car and tested on data from Task plane.

    \setlength{\tabcolsep}{1.9mm}{
    \begin{table*}[ht] 
        \small
        \caption{Comparisons of Balanced-Accuracy ($\%$) of ELIPformer and Comparison Methods in Cross-task Zero-Calibration Experiments (mean $\pm$ standard deviation).}
        \renewcommand\arraystretch{1.22}
        \centering
        \label{comparison experiments cross-task zero-calibration}
        \begin{threeparttable}
        \begin{tabular}{lllllll}
            \toprule[1.2pt]
            \multirow{3}{*}{\textbf{\normalsize Model}}& \multicolumn{6}{c}{\multirow{2}{*}{\textbf{\normalsize Training task $\to$ Test task}}}  \\
            &   &  &  &  &   &\\
            \cmidrule[0.5pt](rl){2-7}
            &  \textbf{car $\to$ plane}  &  \textbf{people $\to$ plane}   &  \textbf{plane $\to$ car}  &  \textbf{people $\to$ car}    & \textbf{plane $\to$ people} & \textbf{car $\to$ people} \\
            \hline	
            HDCA  & $ 82.38\pm5.12^{\star\star\star} $&  $ 
            78.12\pm5.85^{\star\star\star} $ &$81.46 \pm6.03^{\star\star\star} $  &  $ 74.46\pm8.56^{\star\star\star} $ &$75.76 \pm6.75^{\star\star\star} $&$ 72.90\pm7.90^{\star\star\star} $  \\
            MDRM &$ 81.55\pm4.64^{\star\star\star} $&  $78.24 \pm5.33^{\star\star\star} $ &$77.40 \pm 7.92^{\star\star\star}$  &  $73.66 \pm8.89^{\star\star\star} $ &$ 74.95\pm7.38^{\star\star\star} $& $71.91\pm7.95^{\star\star\star} $     \\
            \hline
            EEGNet &$81.89 \pm3.61^{\star\star\star} $  &  $79.10 \pm5.03^{\star\star\star} $  &$82.60 \pm5.41^{\star\star\star} $   &  $77.94 \pm8.39^{\star\star\star} $  &$ 75.16\pm7.61^{\star\star\star} $ &$72.90\pm7.16^{\star\star\star} $      \\
            LeeNet &$ 83.80\pm4.63^{\star\star\star} $ &  $80.05 \pm5.33^{\star\star\star} $   &$83.70 \pm5.09^{\star\star\star} $  &  $ 76.25\pm8.64^{\star\star\star} $ &$ 75.46\pm7.32^{\star\star\star} $&$71.44 \pm8.93^{\star\star\star} $    \\
            PLNet &$79.74 \pm4.56^{\star\star\star} $ &  $ 76.24\pm8.25^{\star\star\star} $ &$80.15 \pm5.69^{\star\star\star} $ &  $74.53 \pm7.86^{\star\star\star} $   &$73.88 \pm7.15^{\star\star\star} $ & $68.55 \pm8.13^{\star\star\star} $  \\
            PPNN& $84.32 \pm4.32^{\star\star\star} $  &  $82.69 \pm4.63^{\star\star\star} $  &$83.31 \pm5.94^{\star\star\star} $  &  $78.59 \pm8.54^{\star\star\star} $  &$76.22 \pm7.10^{\star\star\star} $  &$71.46 \pm8.19^{\star\star\star} $ \\
            HCANN & $ 84.08 \pm3.08 ^{\star\star\star} $&  $ 81.85 \pm4.46 ^{\star\star\star} $ &$ 83.62\pm5.73 ^{\star\star\star} $  &  $77.22 \pm9.77 ^{\star\star\star} $  &$ 75.89\pm8.10 ^{\star\star\star} $ &$ 74.14\pm7.89 ^{\star\star\star} $  \\
            \hline
            HSLT& $ 84.59 \pm3.89 ^{\star\star\star} $   &  $  75.28\pm6.78 ^{\star\star\star} $  & $82.95 \pm5.53^{\star\star\star} $  &  $75.22 \pm 8.74^{\star\star\star} $  &$73.21 \pm7.55 ^{\star\star\star} $ &$ 72.08\pm7.93 ^{\star\star\star} $  \\
            TCN-T & $  85.39\pm3.65 ^{\star\star\star} $ &  $ 82.56 \pm5.10 ^{\star\star\star} $  &$ 83.63\pm5.24 ^{\star\star\star} $  &  $78.26 \pm8.69 ^{\star\star\star} $&$77.40 \pm 6.45^{\star\star\star} $  &$ 75.64\pm7.01^{\star\star\star} $ \\
            TFF-Former & $85.45  \pm 3.38^{\star\star\star} $&  $ 82.95 \pm4.81 ^{\star\star\star} $ &$83.53 \pm6.25 ^{\star\star\star} $  &  $ 79.01\pm8.59 ^{\star\star\star} $  &$ 78.31\pm7.30 ^{\star\star\star} $ &$77.36 \pm6.45 ^{\star\star\star} $  \\
            EEG-conformer & $ 83.69 \pm 3.72 ^{\star\star\star} $&  $  81.57\pm4.59 ^{\star\star\star} $ &$ 82.07\pm5.99 ^{\star\star\star} $  &  $77.70 \pm8.68 ^{\star\star\star} $  &$75.53 \pm7.83 ^{\star\star\star} $ &$74.00 \pm7.20 ^{\star\star\star} $  \\
            \hline
            EEG baseline & $ 84.82\pm3.29^{\star\star\star} $   &  $81.91 \pm 5.02^{\star\star\star}$ & $84.21 \pm4.88^{\star\star} $ &  $79.20 \pm 7.88^{\star\star\star}$ &$ 77.20\pm6.98^{\star\star\star} $  & $ 75.23\pm7.23^{\star\star\star} $     \\
            ELIPformer   &$\mathbf{89.05 \pm2.03}$  &  $\mathbf{89.39 \pm1.74} $& $ \mathbf{86.75\pm3.74} $ &  $\mathbf{85.93 \pm 4.39}$ &$ \mathbf{83.90\pm3.39} $& $\mathbf{82.93 \pm 4.03}$ \\
            \bottomrule[1.2pt]
        \end{tabular}
        \begin{tablenotes}
            \item \footnotesize The asterisks in the table indicate a significant difference between ELIPformer and the comparison methods by paired t-tests ($^{\star} p<0.05,^{\star\star} p<0.01,^{\star\star\star} p<0.001$). The best results are highlighted in bold.
        \end{tablenotes}
        \end{threeparttable}
    \end{table*}}

    \setlength{\tabcolsep}{0.6mm}{
		\begin{table*}[ht]
        \small
        \caption{Balanced-Accuracy ($\%$) of Ablation Experiments (mean $\pm$ standard deviation).}
        \renewcommand\arraystretch{1.22}
        \centering
        \label{ablation study}
        \begin{tabular}{ccccclllllll}
            \toprule[1.2pt]
            \multirow{4}{*}{\textbf{\normalsize Model}}& \multicolumn{4}{c}{\multirow{2}{*}{\large \textbf{\normalsize Ablation Module}}}   &  \multicolumn{6}{c}{\multirow{2}{*}{\textbf{\normalsize Training task $\to$ Test task}}} \\
            &\multicolumn{4}{c}{}   &  \multicolumn{6}{c}{} \\
            \cmidrule[0.5pt](rl){2-5}
            \cmidrule[0.5pt](rl){6-11}
             &\multirow{2}{*}{\textbf{\makecell[c]{EEG}}} & \multirow{2}{*}{\textbf{\makecell[c]{Img}}} & \multirow{2}{*}{\textbf{\makecell[c]{PE}}} & \multirow{2}{*}{\textbf{\makecell[c]{Bi-att}}}   &\makecell[c]{\multirow{2}{*}{\textbf{car $\to$ plane}}}&\makecell[c]{\multirow{2}{*}{\textbf{people $\to$ plane}}}&\makecell[c]{\multirow{2}{*}{\textbf{plane $\to$ car}}}&\makecell[c]{\multirow{2}{*}{\textbf{people $\to$ car}}} & \makecell[c]{\multirow{2}{*}{\textbf{plane $\to$ people}}}& \makecell[c]{\multirow{2}{*}{\textbf{car $\to$ people}}}      \\
            &  &   &   &   &   &  &   &  &  &   &  \\
            \hline	
            M1&$\checkmark$ &--  & --& --&$ 84.82\pm3.29^{\star\star\star} $  &$81.91 \pm 5.02^{\star\star\star}$ & $84.21 \pm4.88^{\star\star\star} $  &$79.20 \pm 7.88^{\star\star\star}$  &$ 77.20\pm6.98^{\star\star\star} $  & $ 75.23\pm7.23^{\star\star\star} $     \\
            M2 &$\checkmark$ &$\checkmark$ & -- &--  & $75.74\pm6.13^{\star\star\star} $ &$62.37\pm5.57^{\star\star\star} $  & $72.56 \pm5.12^{\star\star\star} $ &$ 65.45\pm6.98^{\star\star\star} $  &$61.58 \pm4.24^{\star\star\star} $&$ 70.16\pm6.17^{\star\star\star} $    \\
            M3 &-- &$\checkmark$ & $\checkmark$ &--  & $86.85\pm0.10^{\star\star\star} $ &$86.78\pm0.10^{\star\star\star} $  & $80.49 \pm0.35^{\star\star\star} $ &$ 79.44\pm0.32^{\star\star\star} $  &$81.91 \pm0.04^{\star\star\star} $&$ 81.75\pm0.12 $    \\
            M4 & $\checkmark$& $\checkmark$&$\checkmark$ &--  & $87.65 \pm2.28^{\star\star\star} $ &$ 87.58\pm2.05^{\star\star\star} $ &$84.82 \pm3.60^{\star\star\star} $  & $ 84.76\pm5.31^{\star} $&$81.47 \pm3.60^{\star\star\star} $&  $80.82\pm5.02^{\star\star\star} $       \\
            M5 &$\checkmark$ & $\checkmark$ &$\checkmark$&  $\checkmark$ &$\mathbf{89.05 \pm2.03}$ &$\mathbf{89.39 \pm1.74} $   &$ \mathbf{86.75\pm3.74} $& $\mathbf{85.93 \pm 4.39}$& $ \mathbf{83.90\pm3.39} $ & $\mathbf{82.93 \pm 4.03}$      \\
            \bottomrule[1.2pt]
        \end{tabular}
        \begin{tablenotes} 
            \item \footnotesize ``--” denotes which module is deleted from our model, and``$\checkmark$” denotes which module is remained.
            \item \footnotesize The asterisks in the table indicate a significant difference between our model and the ablation models by paired t-tests ($^{\star} p<0.05,^{\star\star} p<0.01,^{\star\star\star} p<0.001$). The best results are highlighted in bold.
            \item \footnotesize \emph{Abbreviations.} EEG: EEG baseline model; Img: Stimulus image; PE: Prompt encoder; Bi-att: Bidirectional attention.
      \end{tablenotes}
    \end{table*}}
    
    The two-way ANOVA reveals that there are significant main effects of the method ($p < 0.001$) and cross-task experiment ($p < 0.001$), as well as an interaction effect between factors on classification performance ($p < 0.01$). The post-hoc tests show that the BA of ELIPformer is significantly higher than that of all comparison models in each cross-task experiment (all: $p<0.01$). Therefore, our method can significantly enhance the performance of cross-task zero-calibration RSVP decoding. HDCA, PPNN, and TFF-Former achieve the best performance among the conventional, CNN-based, and Transformer-based comparison methods, respectively. TFF-Former outperforms both HDCA and PPNN in each cross-task experiment. This indicates that the Transformer-based network can improve decoding performance compared to conventional machine learning and CNN-based methods. It is noteworthy that our EEG baseline model performs better than HDCA and PPNN, and achieves comparable performance to TFF-Former in five experiments. Hence, our EEG baseline model is effective for EEG feature extraction. Based on the EEG baseline model, our proposed ELIPformer incorporates language-image prior knowledge and realizes semantic alignment between EEG features and prior knowledge, which achieves the best performance among the comparison methods in all cross-task experiments. These results demonstrate the efficacy of integrating language-image prior knowledge into the EEG decoding model to enhance cross-task RSVP decoding.

\subsection{Ablation Study}
    To evaluate the effect of each part in our proposed method, we conduct an ablation experiment on the six cross-task zero-calibration experiments. First, we assess the performance of the EEG baseline model (M1). Then, we conduct three experiments (M2, M3, and M4) to analyze the efficacy of the prompt encoder and bidirectional attention mechanism. In M2, the EEG baseline is combined with image features extracted by the pre-trained ViT-B/32 in CLIP, where the image features are concatenated with EEG features for classification. In M3, we employ a model that relies only on language-image features extracted from task-specific prompts and stimulus images by the prompt encoder. In M4, we combine the EEG baseline with language-image features extracted by the prompt encoder. The structure of M4 is identical to ELIPformer, except that the bi-attention mechanism in the cross bi-attention module is replaced by the original attention in \cite{tsai2019multimodal}. Finally, we compare the classification performance of the M4 with ELIPformer (M5) which employs the proposed cross bi-attention mechanism to evaluate the effectiveness of the bidirectional attention mechanism. The results are shown in Table \ref{ablation study}. 
    
    The two-way ANOVA indicates significant main effects of the ablation module ($p < 0.001$) and cross-task experiment ($p < 0.001$), as well as an interaction effect between these two factors on classification performance ($p < 0.001$). Post-hoc tests show that our proposed ELIPformer (M5) significantly outperforms all other ablation models in each cross-task experiment (all: $p<0.05$) except for M3 in car$\to$people experiment. However, the performance of M5 tends to be significantly higher than that of M3 in car$\to$people experiment ($p=0.15$). Compared with the EEG baseline model (M1), M2 combines the pre-trained ViT of CLIP and the EEG baseline model (M1), but its performance significantly decreases in each cross-task experiment (all: $p<0.001$). This is attributed to the semantic gap between the image features (representing image categories) extracted by pre-trained ViT and the EEG features (representing task-related information, i.e., target or nontarget) extracted by the EEG decoding model. Notably, the model utilizing only image features extracted by our proposed prompt encoder (M3) performs better than the model that integrates both EEG features and image features extracted by pre-trained ViT (M2), which shows the effectiveness of the prompt encoder in discriminating between target and nontarget using task-specific prompts and stimulus images. M4 integrates the language-image prior knowledge extracted by the prompt encoder into the EEG baseline model, which outperforms M1 in five cross-task experiments (all: $p<0.001$) and M3 in four cross-task experiments (all: $p<0.001$). This indicates that language-image features extracted by the prompt encoder can enhance decoding performance after semantic alignment with EEG features. Finally, our proposed model (M5) incorporating a bidirectional attention mechanism based on M4 leads to a significant improvement in the decoding performance compared to M4 with the original attention mechanism (all: $p<0.05$). These results demonstrate that each component in our model significantly enhances decoding performance.

\subsection{The Effect of Prompt Encoder}
    To illustrate the effect of prompt encoder in extracting task-specific language-image prior knowledge, t-distributed Stochastic Neighbor Embedding (tSNE) \cite{van2008visualizing} is applied to project the image class tokens extracted by the pre-trained ViT of CLIP and prompt encoder into two dimensions, respectively. We randomly select fifty target images and fifty nontarget images from each task. 

    \begin{figure}[ht] 
        \centering
        \subfigure[]{
            \centering
            \includegraphics[width=0.41\linewidth]{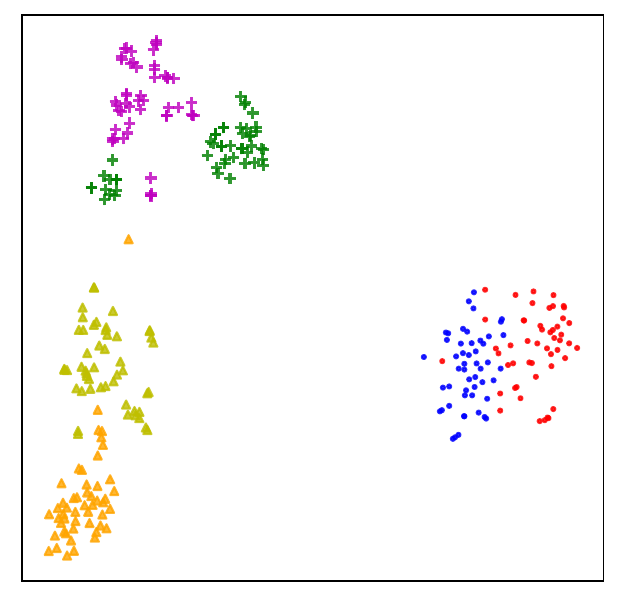}
        }
        \subfigure[]{
                \centering
                \includegraphics[width=0.41\linewidth]{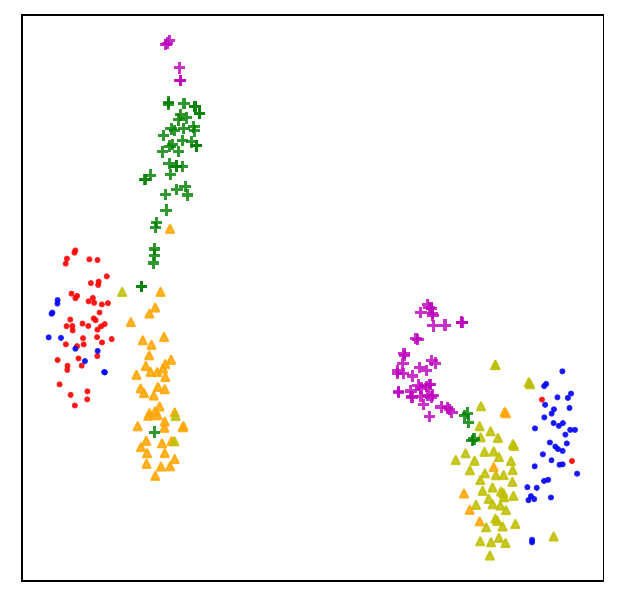}
        }\\
        \subfigure{
                \centering
                \includegraphics[width=0.95\linewidth]{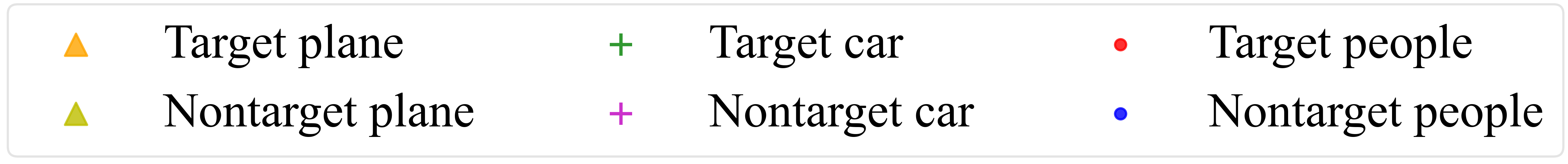}
        }
        \centering
        \caption{The t-SNE visualization of the class token features extracted by (a) the pre-trained ViT of CLIP and (b) our proposed prompt encoder from stimulus images in three tasks.}
        \label{tSNE}
    \end{figure}
    
    As shown in Fig. \ref{tSNE}(a), the features extracted by the pre-trained ViT of CLIP are tightly clustered together with the same task. In contrast, features extracted by the prompt encoder are more closely linked to target or nontarget and narrow the distance of features in the same class across different tasks (see Fig. \ref{tSNE}(b)). This can be attributed to the fact that the prompt encoder incorporates task-related semantic embedding into image tokens before feeding them into the Transformer layers, thereby facilitating the extraction of discriminative language-image features between task-specific target and nontarget.
    
\subsection{The Effect of Cross Bi-attention Module}
\begin{figure}[ht] 
        \centering
        \includegraphics[width=0.85\linewidth]{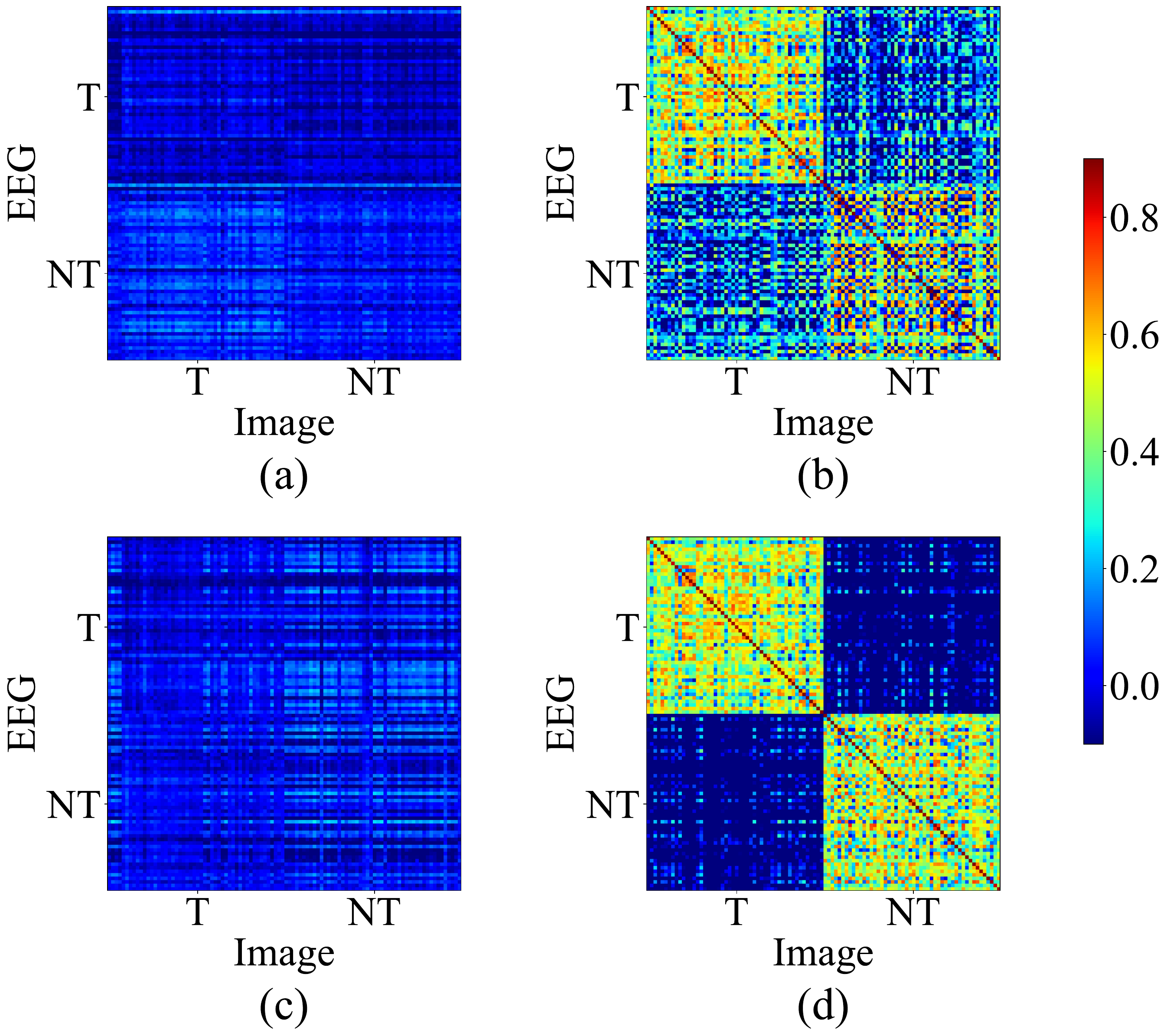}
        \caption{The cosine similarity between EEG features and language-image features before and after the interaction process using the cross-attention mechanism and cross bi-attention mechanism. (a), (b) are the similarity of features before and after interaction using cross-attention module with original attention mechanism in \cite{tsai2019multimodal}, respectively. (c), (d) are the similarity of features before and after interaction using cross bi-attention module with bidirectional attention mechanism, respectively. The ``T" and ``NT" denote target and nontarget respectively.}
        \label{similarity}
    \end{figure}

    To evaluate the impact of the bidirectional attention mechanism on feature interaction, we visualize the cosine similarity between EEG and language-image features before and after interaction using the conventional attention mechanism \cite{tsai2019multimodal} and the proposed cross bi-attention mechanism. Two models are trained: one with a cross-attention module using the original attention mechanism and the other with the cross bi-attention mechanism. Fifty target and fifty nontarget samples from the same subject are randomly selected for testing. Heatmaps visualize the cosine similarity between the flattened EEG tokens and image tokens before and after interaction using the cross-attention module. The results from the car$\to$plane experiment are shown in Fig. \ref{similarity}.

    From Fig. \ref{similarity}(a) to (b) and (c) to (d), the similarity between EEG and language-image features within both target and non-target classes increases after interaction. This demonstrates that both attention mechanisms effectively align EEG and language-image features. Compared to the original attention mechanism (Fig. \ref{similarity}(b)), the bi-attention mechanism (Fig. \ref{similarity}(d)) not only improves similarity within the same class but also reduces similarity across different classes. This highlights the bi-attention mechanism's superior ability to align features and decrease the False Positive Rate (FPR). Additional ablation experiments confirm that the FPR of ELIPformer using the bi-attention mechanism is significantly lower than that using the original attention mechanism in all six cross-task experiments (all: $p < 0.001$).

\subsection{Evaluation of Two Training Tasks}
    In each experiment presented in Table \ref{comparison experiments cross-task zero-calibration}, we utilize a single task as the training set to train ELIPformer. In this part, we conduct experiments where the training data is composed of multiple tasks. We employ each task as the test set, while the remaining two tasks are utilized as the training set. The results are presented in Table \ref{two tasks training}. The one-way repeated ANOVA demonstrates the significant effect of the training tasks on performance for each test task (all: $p<0.001$). Moreover, training the model with data from two tasks significantly improves performance compared to using data from a single task (all: $p<0.001$). These results demonstrate that our model can be trained using EEG signals and stimulus images from different existing RSVP tasks simultaneously, and can effectively use data from more subjects to enhance decoding performance. This is because the ELIPformer can map the features of images from diverse tasks into the same semantic space and establish semantic alignment between EEG and image features with the guidance of prompts.

    \setlength{\tabcolsep}{0.6mm}{
		\begin{table}[ht] 
            \begin{threeparttable}
            \footnotesize 
            \renewcommand\arraystretch{1.27}
            \centering
            \caption{Balanced-Accuracy ($\%$) of ELIPformer Training on Two Tasks (mean $\pm$ standard deviation).}
            \begin{tabular}{llllll}
                \toprule[1.2pt]
                \textbf{Train}  &  \centering{ \textbf{Test (plane)}}   &  \textbf{Train}  &  \textbf{Test (car)}    & \textbf{Train}  &  \textbf{Test (people)} \\
                \hline
				car  &  $89.05\pm2.03^{\star\star\star}$   &  plane  &  $ 86.75\pm3.74^{\star\star\star} $    & plane  &  $ 83.90\pm3.39^{\star\star\star} $ \\
                people  &  $89.39 \pm1.74^{\star\star\star} $   &  people  &  $85.93 \pm 4.39^{\star\star\star}$    & car  &  $82.93 \pm 4.03^{\star\star\star}$ \\
                \hline	
                both  &  $\mathbf{90.40 \pm 1.65}$   &  both  &  $\mathbf{88.84 \pm 3.45}$    & both  & $\mathbf{85.22 \pm 3.16}$  \\
                \bottomrule[1.2pt]
            \end{tabular}
            \begin{tablenotes}
				\item \footnotesize The asterisks in the table indicate a significant difference between the model trained on two tasks and the model trained on one task by paired t-tests ($^{\star} p<0.05,^{\star\star} p<0.01,^{\star\star\star} p<0.001$). The best results are highlighted in bold.
		  \end{tablenotes}
            \label{two tasks training}
            \end{threeparttable}
	\end{table}} 

    \begin{figure*}[ht] 
        \centering
        \subfigure[]{
            \centering
        \includegraphics[width=0.23\linewidth]{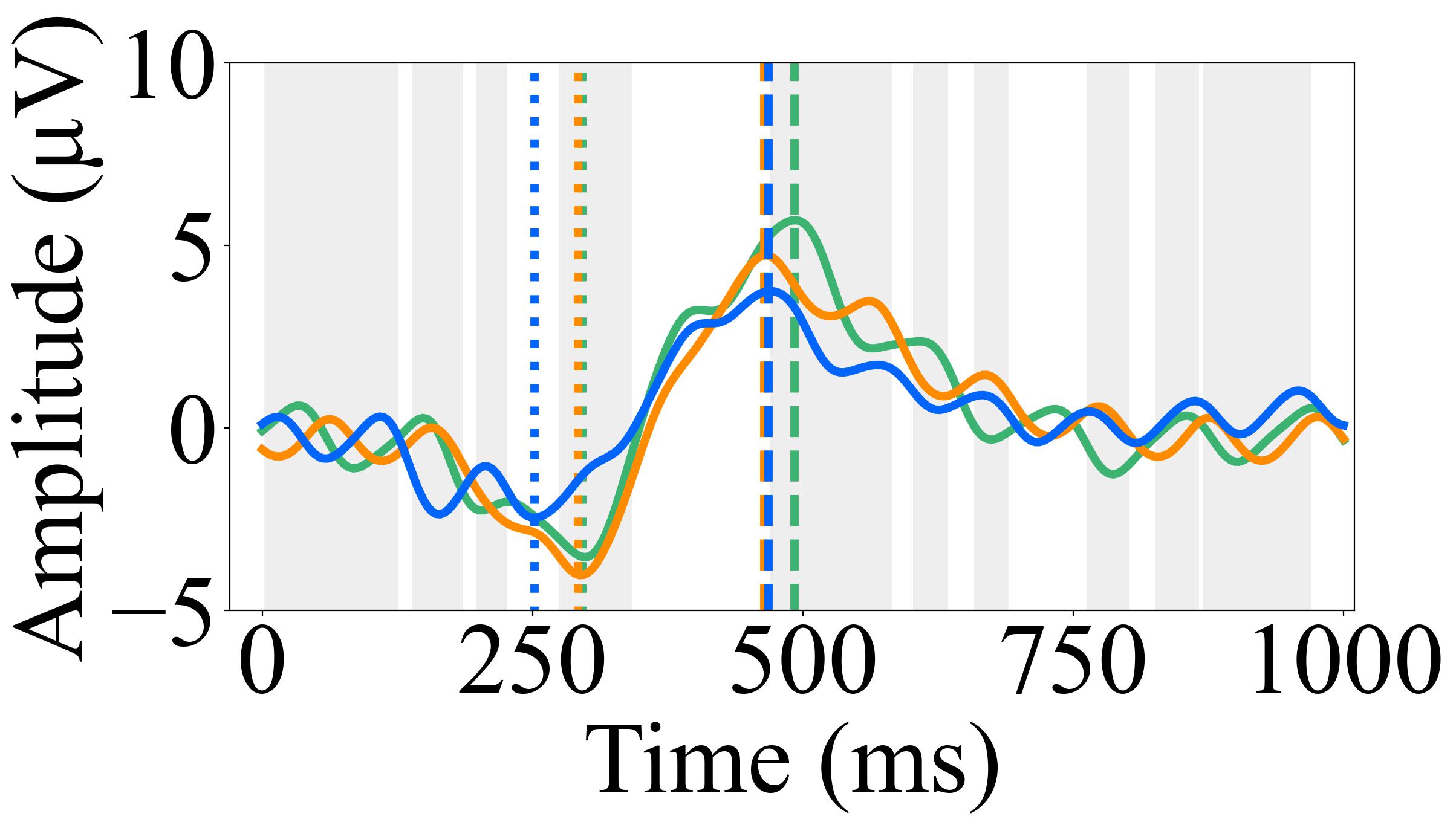}
        }
        \subfigure[]{
            \centering
        \includegraphics[width=0.23\linewidth]{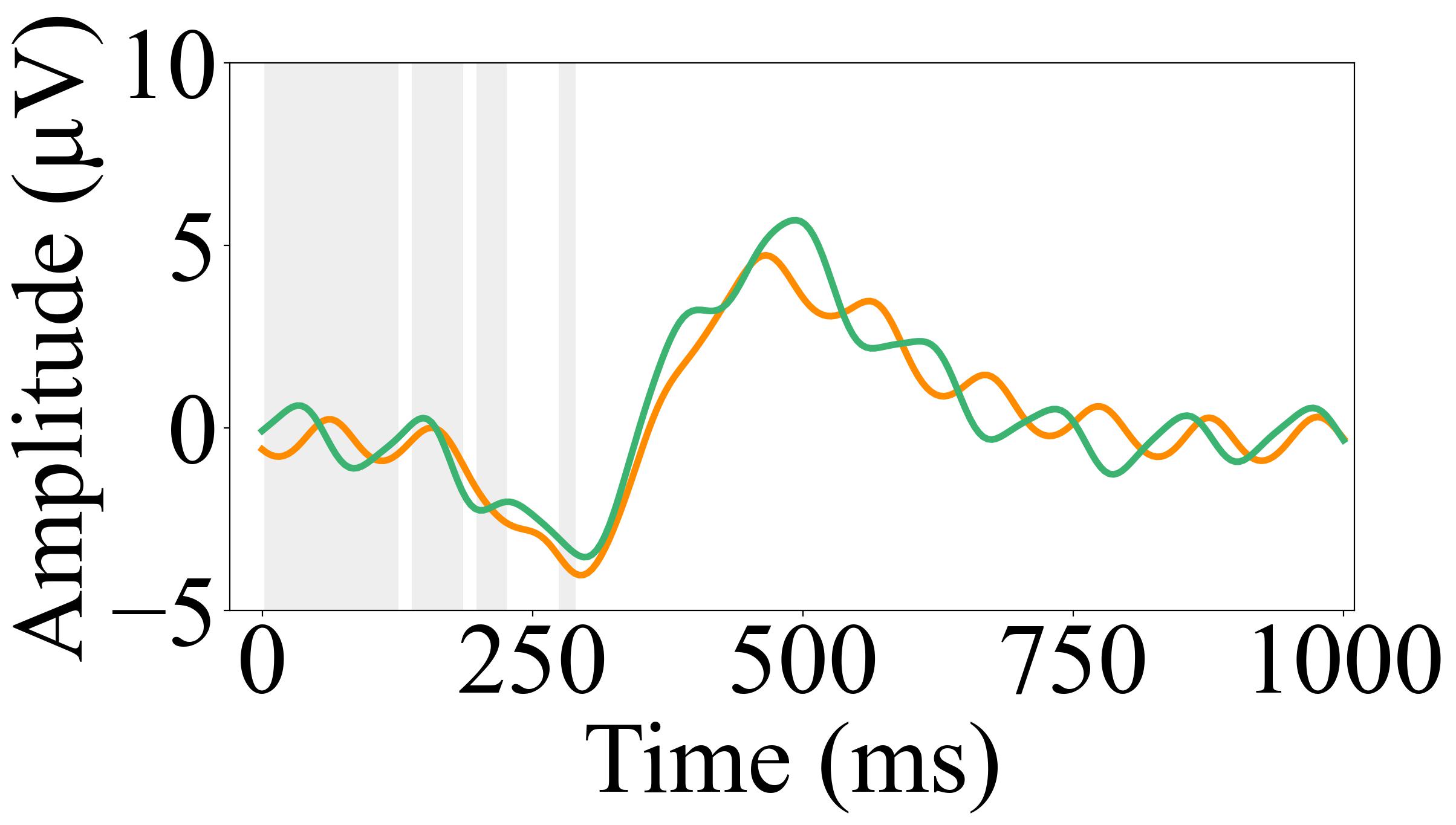}
        }
        \subfigure[]{
            \centering
        \includegraphics[width=0.23\linewidth]{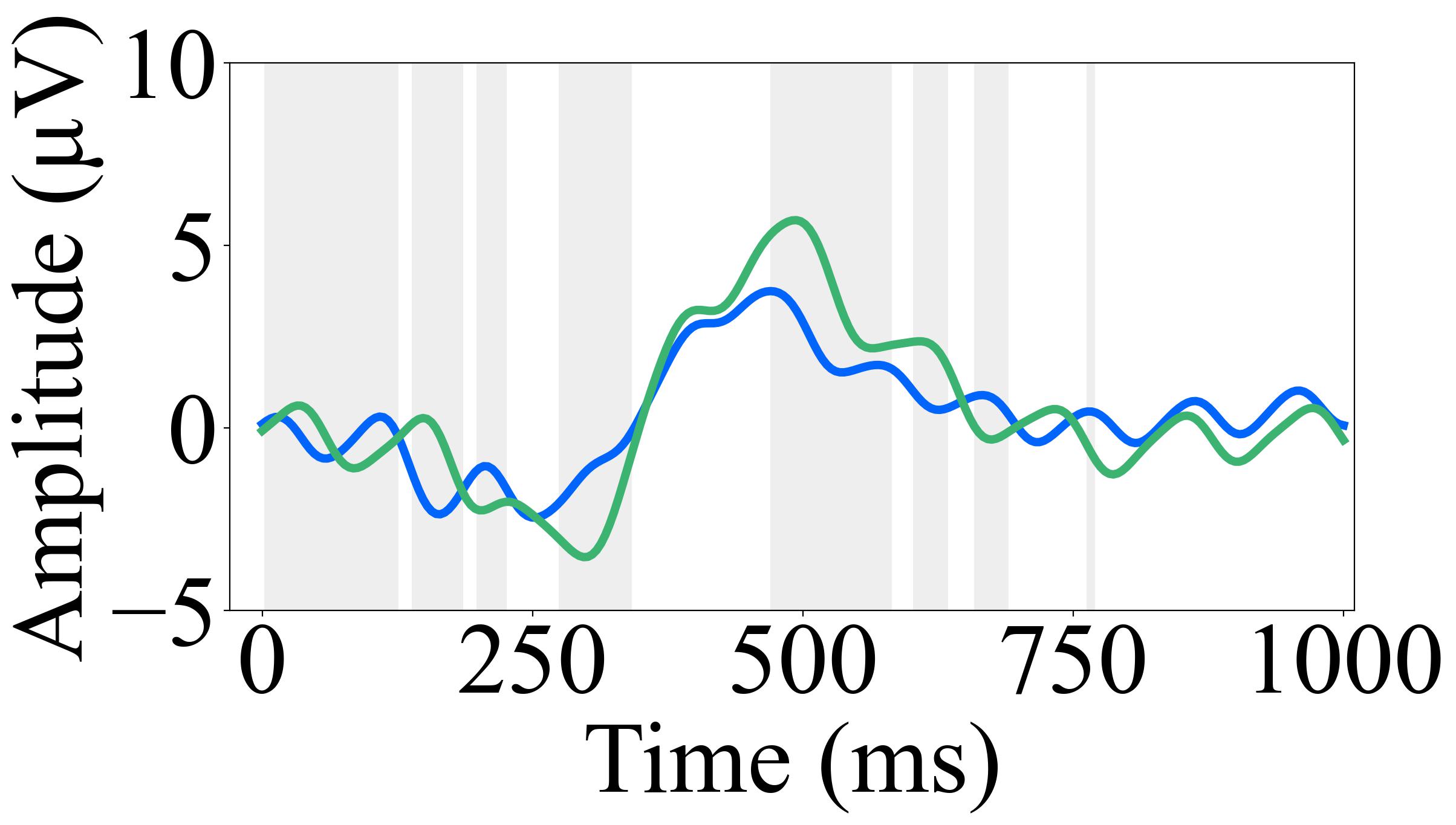}
        }
        \subfigure[]{
            \centering
        \includegraphics[width=0.23\linewidth]{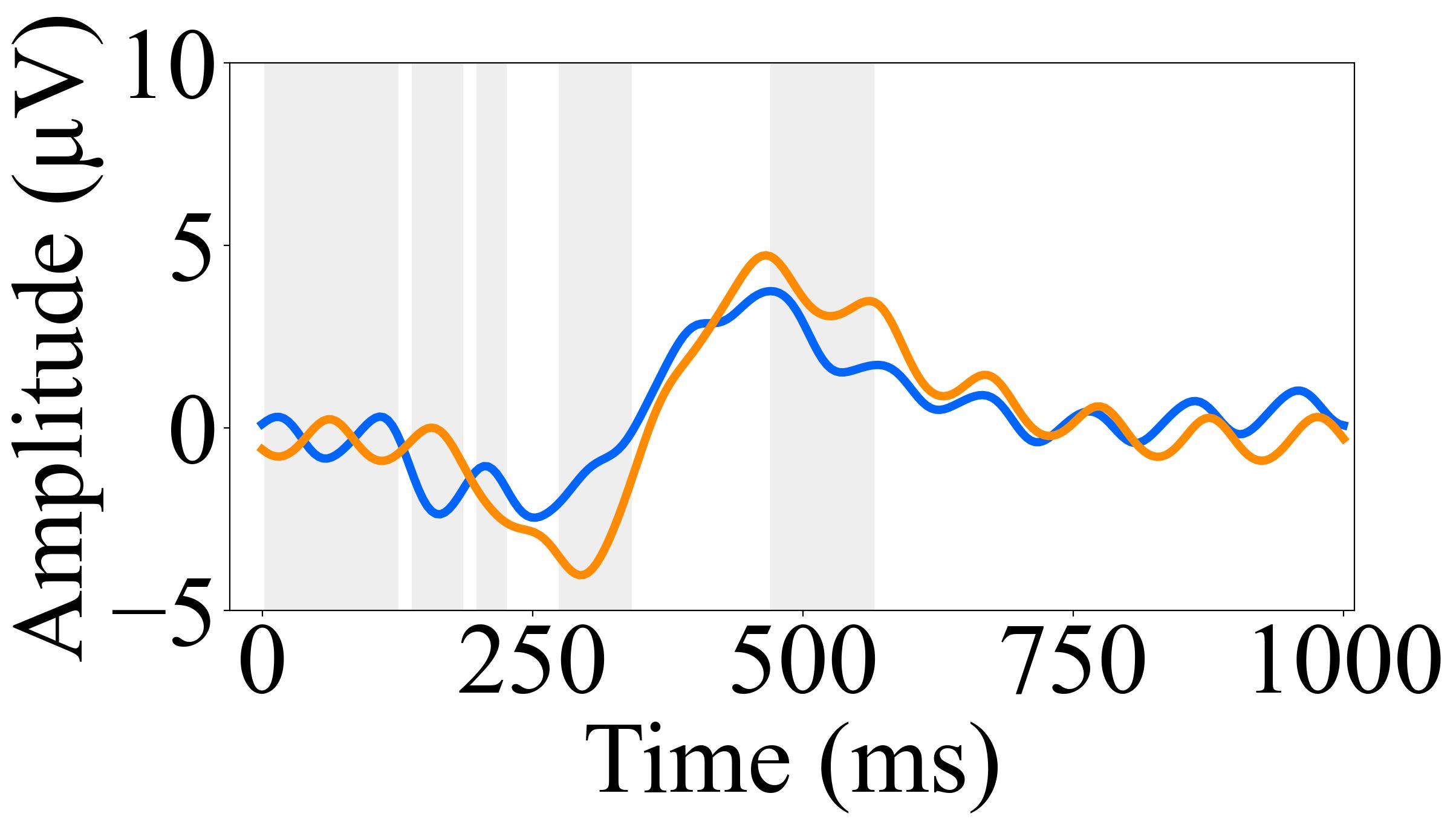}
        }
        \subfigure{
            \centering
        \includegraphics[width=0.6\linewidth]{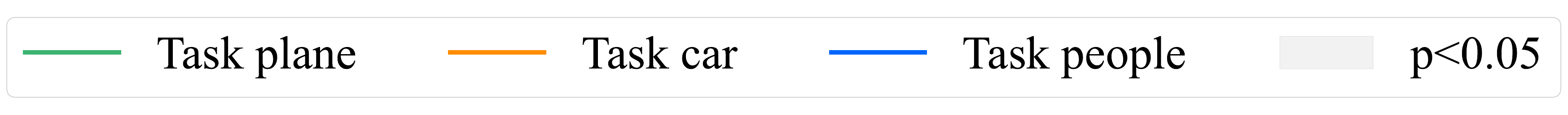}
        }\vspace{-0.2cm}
        \caption{The visualization of ERP responses. (a) The one-way ANOVA analysis of grand-average target EEG waveforms from Oz channels of three tasks. (b), (c), (d) The post-hoc pairwise tests of grand-average target EEG waveforms from Oz channels. The shaded regions represent significant differences ($p<0.05$) among different tasks at that moment.}
        \label{EEG analysis}
    \end{figure*}

\subsection{Analysis of EEG Responses}  
    To investigate differences in EEG responses across tasks, we plot the grand-averaged ERP waveforms of the representative Oz channel in RSVP tasks \cite{mao2023cross,zhou2024rsvp} and EEG topographies across all subjects for Task plane, Task car, and Task people, respectively (see Fig. \ref{EEG analysis}). As shown in Fig. \ref{EEG analysis}(a), the typical N200 and P300 components are visible for all three tasks. One-way ANOVA reveals a significant main effect of the task on ERP waveform amplitude, particularly in the 280-340 ms and 470-580 ms intervals where the N200 and P300 responses appear, respectively \cite{patel2005characterization}. In Fig. \ref{EEG analysis}(b)-(d), post-hoc tests indicate significant differences ($p<0.05$) in many moments in pairwise comparisons among the three tasks. Specifically, all pairwise comparisons show significant differences ($p<0.05$) in the P300 response period, while there are significant differences in the N200 period between Task people and Task plane, and between Task people and Task car.

    \begin{figure}[ht] 
        \centering
        \subfigure{
                \centering
                \includegraphics[width=0.85\linewidth]{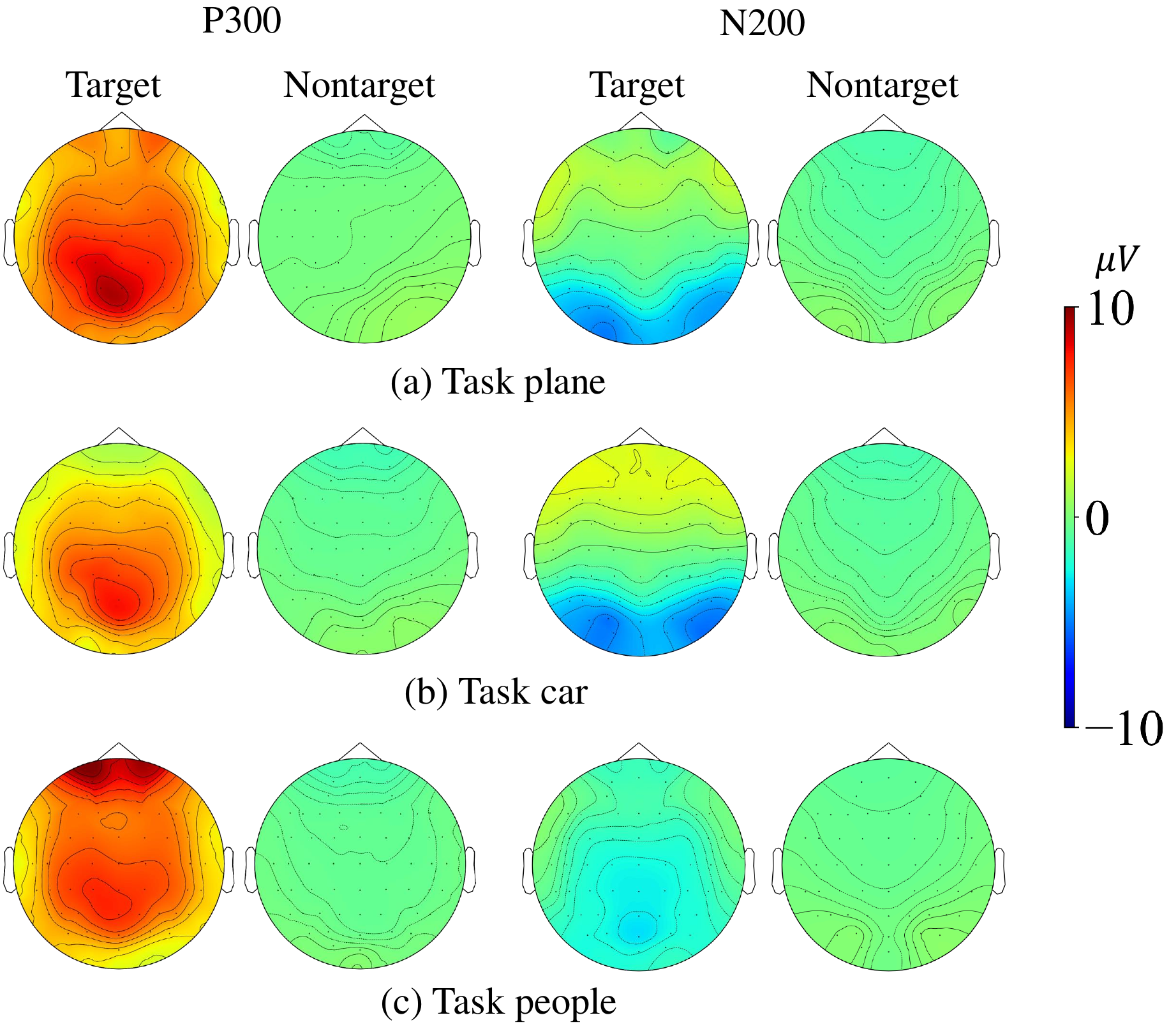}
        }
        \caption{The grand-average target and nontarget EEG topographies at the peak moment of P300 and N200 components across all subjects in (a) Task plane, (b) Task car, and (c) Task people, respectively.}
        \label{EEG topographic map}
    \end{figure}
    
    The target and nontarget EEG topographies at the peak moments of the P300 and N200 responses for the Oz channel across three tasks are shown in Fig. \ref{EEG topographic map}(a)-(c). For target EEG responses, the P300 and N200 responses are mainly concentrated in the parietal lobe and occipital lobe. The one-way ANOVA reveals significant differences in the P300 responses across the three tasks in six channels located in the occipital lobe which is associated with visual processing (all: $ p<0.05$). Significant differences are also observed in the N200 responses across tasks in channels located in both the occipital and parietal lobes (all: $p<0.05$), with the N200 responses of Task people in the parietal lobe being stronger than those of Task plane and Task car. These variations in EEG signals across tasks pose challenges for cross-task RSVP decoding.

\subsection{Visualization of Data Discriminative Region}
    The Gradient-weighted Class Activation Mapping (Grad-CAM) \cite{selvaraju2017grad} is widely used to generate class activation maps that highlight discriminative regions for the class of interest. It uses the class-specific gradient information without modifying the model structure. In this study, we utilize the Grad-CAM on the outputs of the cross bi-attention module to generate the activation value for corresponding categories. For each task, we input a target and nontarget sample from the same subject into the trained model to investigate the regions of input data the model focuses on.

    \begin{figure*}[ht] 
        \centering
        \includegraphics[width=\linewidth]{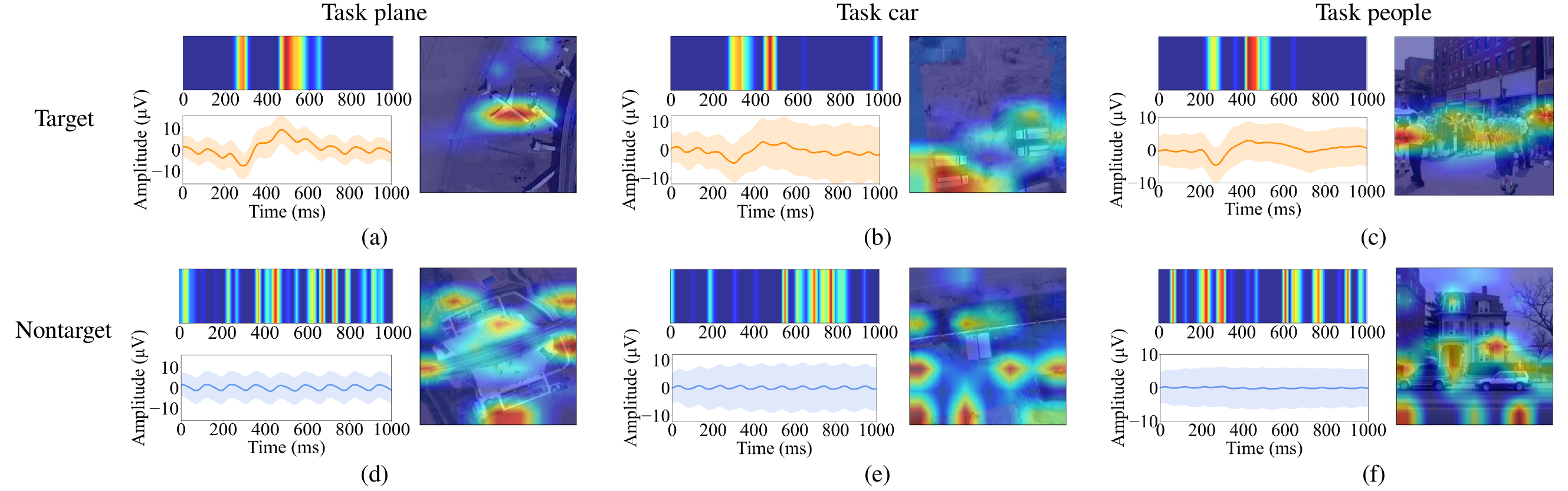}\vspace{-0.15cm}
        \caption{The Grad-CAM visualization of EEG signals and the corresponding stimulus images generated using ELIPformer in three tasks. Each subfigure contains three components. The upper left are the activation values of EEG to the classification results, and the lower left are the average waveforms at the Oz channel of the corresponding subjects. On the right are the activation values of stimulus images to the classification results overlaid on the corresponding images.}
        \label{Grad cam}
    \end{figure*}

    In Fig. \ref{Grad cam}, each subfigure contains three elements: the activation values of EEG to the classification results, the activation values of stimulus images to the classification results, and averaged waveforms at the Oz channel of the corresponding subject. The left side shows activation values from EEG tokens output by the cross bi-attention module, reflecting the importance of EEG temporal information in classification. The right displays the activation values overlaid on the corresponding image, which are obtained from the image tokens output by the cross bi-attention module and represent the importance of image regions to classification. High activation values are observed for target EEG signals in the 200–300 ms and 400–500 ms intervals, corresponding to the N200 and P300 responses (see Fig. \ref{Grad cam}(a)-(c)). For target images, high activation values are concentrated on the regions containing the targets of each task (i.e. plane, car, and people respectively). In contrast, the regions obtaining high activation values are scattered in both nontarget EEG signals and images (see Fig. \ref{Grad cam}(d)-(f)). These results suggest that our model can effectively extract and employ task-related features from both EEG signals and stimulus images for the final classification.

\subsection{Consideration of Cross-task Decoding Performance}
    In Table \ref{comparison experiments cross-task zero-calibration}, the performance of ELIPformer and other decoding models tested on the task people is lower than that of those models tested on the task plane and task car. This discrepancy can be attributed to two potential reasons. 
    
    On the one hand, in Fig. \ref{EEG analysis}(b), the grand-average target EEG waveforms from Oz channels in the Task plane are significantly different from those in the Task car at 21.6\% of the moments. Conversely, Fig. \ref{EEG analysis}(c) and (d) show significant differences between Task people and Task plane at 45.2\% of the moments, and between Task car and Task people at 36.4\% of the moments. Additionally, in Fig. \ref{EEG topographic map}, compared to the other two tasks, the EEG topographic map of Task people shows stronger responses in the parietal lobe and weaker responses in the occipital lobe at the peak moment of N200. These results suggest that the difference between EEG signals of Task people and the other two tasks is greater than the differences between those of Task plane and Task car, which leads to decreased performance of the decoding model in plane$\to$people and car$\to$people experiments compared to plane$\to$car and car$\to$plane. This reason is also supported by the lower performance of EEG decoding models in people$\to$plane and people$\to$car compared to car$\to$plane and plane$\to$car, respectively.
    
    \begin{figure}[ht] 
        \centering
        \subfigure{
                \centering
                \includegraphics[width=0.65\linewidth]{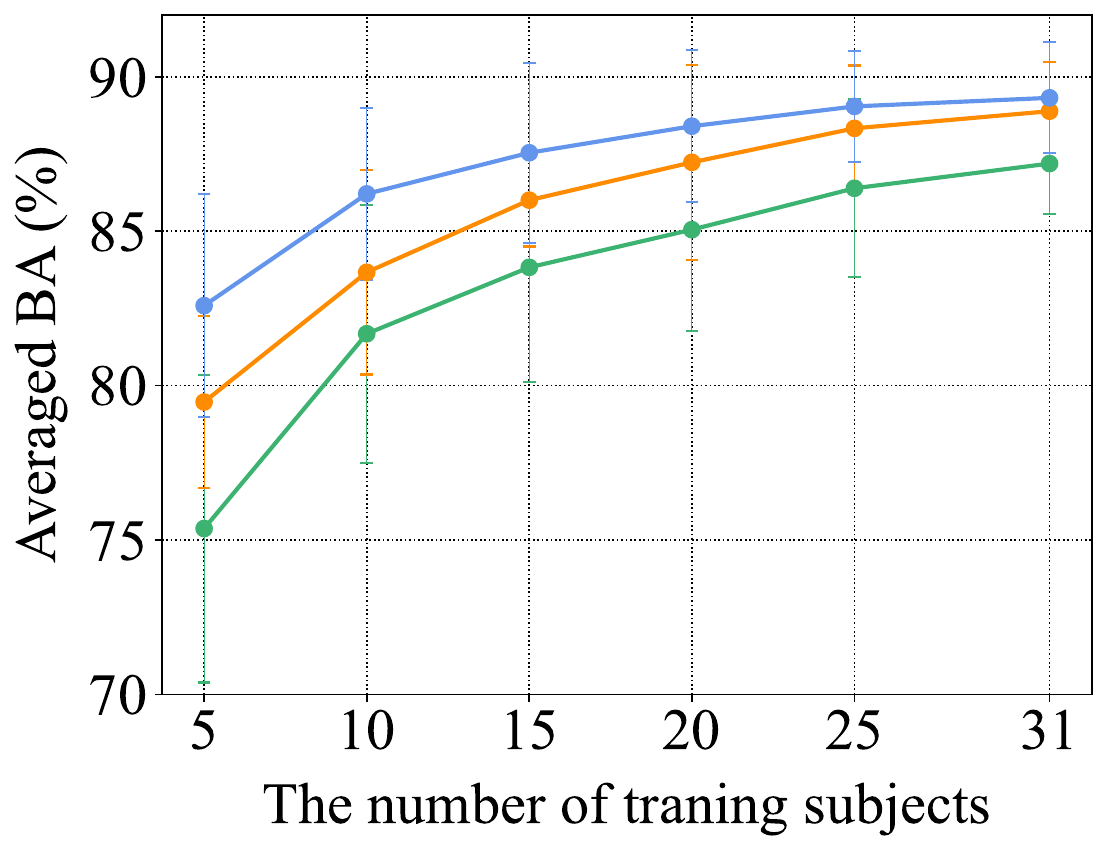}
        }
        \subfigure{
                \centering
                \includegraphics[width=0.7\linewidth]{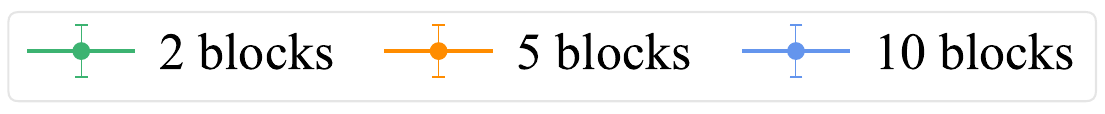}
        }\vspace{-0.2cm}
        \caption{The changes in balanced-accuracy of ELIPformer with different numbers of training subjects each containing 2, 5, and 10 blocks of data, respectively.}
        \label{training data volume}
    \end{figure} 
    
    On the other hand, when two tasks alternate as the training and test tasks in cross-task experiments, the diversity of training subjects could influence the performance (plane$\to$people $<$ people$\to$plane; car$\to$people $<$ people$\to$car). Task people includes 31 subjects, while Task plane and Task car each include 20 subjects. To verify this point, we evaluate the ELIPformer performance in the people$\to$plane experiment with varying numbers of training subjects each using 2, 5, and 10 blocks of data, respectively (see Fig. \ref{training data volume}). The BA of the model improves significantly with each increase of five training subjects for 2 and 5 training blocks (all: $p<0.05$). For 10 training blocks, the BA of the model also displays a significant enhancement from 5 subjects to 25 subjects, and the model trained on 31 subjects achieves significantly higher BA compared to the model trained on 20 subjects (all: $p<0.05$). These results suggest that augmenting the subject diversity can effectively enhance the decoding performance. Furthermore, we observe that when two training sets possess an equal number of samples, the model trained on the dataset from more subjects performs significantly better than that trained on a training set that contains more blocks from fewer subjects (all: $p<0.05$). For instance, the model trained with 25 subjects each containing 2 blocks of data achieves a significantly higher performance (86.39\%) in contrast to the model trained with 10 subjects each containing 5 blocks of data (83.67\%). These results further support the importance of subject diversity in improving model performance.
    
\subsection{Limitation and Future Work}
    Our proposed EEG decoding model extracts language-image features from task-specific prompts and stimulus images to enhance cross-task zero-calibration RSVP-EEG decoding. However, the approach has several limitations. The prompts in the model need to be artificially designed. In the future, the prompt learning techniques from natural language processing can be used to learn a general template. Also, our RSVP experiments only consider the typical RSVP presentation speed of 10Hz, and future work will investigate the impact of different RSVP speeds.
    
\section{Conclusion}
    In this study, we propose an EEG with language-image prior fusion Transformer (ELIPformer) to enhance cross-task zero-calibration RSVP decoding. The model utilizes a prompt encoder to extract language-image features from task-specific prompts and stimulus images as prior knowledge and employs a cross bi-attention module to achieve effective feature fusion and semantic alignment between EEG and language-image features. We design three distinct target image retrieval tasks and construct an open-source cross-task RSVP dataset comprising EEG signals and corresponding stimulus images. Experimental results demonstrate that our method outperforms previous approaches in cross-task RSVP decoding across all three tasks. These results suggest that the RSVP-BCI system based on our approach enables rapid deployment and efficient target retrieval in diverse scenarios, which represents an important step towards the practical application of RSVP-BCI systems.

%References
\bibliographystyle{bare_jrnl_compsoc}
\bibliography{bare_jrnl_compsoc}

\vfill

\end{document}